\documentclass[twocolumn,preprintnumbers,amsmath,amssymb]{revtex4}  

\usepackage{amsfonts}
\usepackage[T1]{fontenc}
\usepackage{amsmath,amsbsy,amssymb,graphicx}
\usepackage{times}
\usepackage{amsmath}
\usepackage{amssymb}
\usepackage{graphicx}
\usepackage{color}

\begin{document}

\title{Graphane with carbon dimer defects: Robust in-gap states and a scalable two-dimensional platform for quantum computation}

\author{Lei Hao$^{1,2}$, Hong-Yan Lu$^{3,4}$, and C. S. Ting$^{2}$}
 \address{$^1$School of Physics, Southeast University, Nanjing 211189, China \\
$^2$Texas Center for Superconductivity and Department of Physics,
University of Houston, Houston, Texas 77204, USA \\
$^3$School of Physics and Physical Engineering, Qufu Normal University, Qufu 273165, China \\
$^4$School of Physics and Electronic Information, Huaibei Normal University, Huaibei 235000, China}

\date{\today}

\begin{abstract}
We study the energy level structures of the defective graphane lattice, where a carbon dimer defect is created by removing the hydrogen atoms on two nearest-neighbor carbon sites. Robust defect states emerge inside the bulk insulating gap of graphane. While for the stoichiometric half-filled system there are two doubly degenerate defect levels, there are four nondegenerate and spin-polarized in-gap defect levels in the system with one electron less than half filling. A universal set of quantum gates can be realized in the defective graphane lattice, by triggering resonant transitions among the defect states via optical pulses and \emph{ac} magnetic fields. The sizable energy separation between the occupied and the empty in-gap states enables precise control at room temperature. The spatial locality of the in-gap states implies a qubit network of extremely high areal density. Based on these results, we propose that graphane as a unique platform could be used to construct the future all-purpose quantum computers.
\end{abstract}


\maketitle

\section{Introduction}

Since the discovery of quantum algorithms that outperform all existing classical algorithms \cite{deutsch85,deutsch92,shor94,grover96}, there have been growing interest in building a real-world quantum computer. This desire is further fostered by the suggestion of Feynman that quantum computers can simulate complex quantum mechanical systems more efficiently than classical computers \cite{feynman82}. At present, although proof-of-principle demonstrations of quantum computation exist \cite{chuang98,debnath16} and the quantum supremacy seems to be on the horizon \cite{preskill12,boixo18}, a practical all-purpose quantum computer is at most in its early infancy. One major challenge is to find a unique platform that can play the role of silicon wafer in classical (digital) computers. In this respect, a two-dimensional (2D) solid-state system is highly advantageous. Existing examples of such platforms include the semiconductor quantum dots \cite{loss98,trauzettel07,veldhorst17} and the Josephson junction arrays \cite{makhlin01,you11}. These artificial systems, while highly tunable, have the deficiency that the basic constituents (quantum dots, Josephson junctions) cannot be reproduced exactly, and the properties of different qubits (quantum two-level systems) vary unavoidably from one to another. This shortcoming can lead to considerable cumulative error in large-scale quantum computations. The defect energy levels in a crystalline insulator, like the nitrogen-vacancy centres in diamond \cite{zu14} and $^{31}$P impurities in silicon (the Kane quantum computer \cite{kane98}), are free of this problem. The diamond unfortunately is a three-dimensional system. The Kane quantum computer, being a quasi-2D framework, relies on the artificial engineering on the depth of the $^{31}$P impurities underneath the silicon surface and the distances between individual $^{31}$P impurities. Therefore, a purely 2D solid-state quantum computing network with stably reproducible energy level structure to encode the qubits was lacking and is highly desirable.

Here we show that graphane, the fully hydrogenated graphene, could be an ideal 2D platform for quantum computing. Graphane was theoretically predicted to be an insulator with a huge band gap lying in the range from 3.42 eV to 5.97 eV \cite{sofo07,sluiter03,lebegue09,sahin09}, and was also synthesized experimentally \cite{elias09}. Introducing hydrogen vacancies to graphane leads to in-gap states. Previous works focus on the hydrogen vacancies that break the balance of the two sublattices and induce ferromagnetic polarization \cite{lebegue09,sahin09,sahin10,yang10,berashevich10,islam16,mapasha16,sahin15}. We demonstrate that the simplest hydrogen vacancy configuration keeping the balance of the two sublattices, which is named as a carbon dimer and is created by removing the hydrogen atoms on two nearest-neighbor (NN) carbon sites, gives rise to interesting in-gap defect states. Being intrinsic defects in a crystalline insulator, different carbon dimers have exactly the same in-gap states. In the stoichiometric half-filled system, each carbon dimer gives rise to two doubly-degenerate defect states. By removing one electron from the graphane lattice per carbon dimer, the two-fold degeneracy in the defect energy levels of the half-filled system is removed, and there are four nondegenerate in-gap defect states associated with each carbon dimer. We show that the defect states are tunable by a gate voltage and controllable by external stimuli such as laser pulses and \emph{ac} magnetic fields. In particular, both single-qubit and nontrivial two-qubit unitary operations can be realized on these defect states, thereby enabling universal quantum computation. Based on the analyses, we discuss the feasibility of utilizing the graphane lattice with carbon dimer defects as a 2D solid-state platform for scalable quantum computation.

\section{\label{sec:energetics}Energy structures of graphane with a carbon dimer defect}

We focus on the qualitative changes in the quasiparticle spectrum of graphane, before and after the creation of one carbon dimer. A minimal model for this purpose is constructed by retaining the carbon $p_{z}$ orbitals and the hydrogen $s$ orbitals. We therefore consider a tight-binding model up to next-nearest-neighbor (2NN) hopping of the carbon $p_{z}$ orbitals, supplemented by the local carbon-hydrogen (C-H) bonds, written as
\begin{eqnarray}
&&\hat{H}_{graphane}=t_{NN}\sum\limits_{\langle ij\rangle\sigma}c^{\dagger}_{i\sigma}c_{j\sigma}
+t_{2NN}\sum\limits_{\ll ij\gg\sigma}c^{\dagger}_{i\sigma}c_{j\sigma}     \notag \\
&&+\epsilon_{p}\sum\limits_{i\sigma}c^{\dagger}_{i\sigma}c_{i\sigma}
+\epsilon_{s}\sum\limits_{i\sigma}h^{\dagger}_{i\sigma}h_{i\sigma}
+U\sum\limits_{i}c^{\dagger}_{i\uparrow}c_{i\uparrow}c^{\dagger}_{i\downarrow}c_{i\downarrow}   \notag \\
&&+t_{CH}\sum\limits_{i\sigma}(c^{\dagger}_{i\sigma}h_{i\sigma}+h^{\dagger}_{i\sigma}c_{i\sigma}).
\end{eqnarray}
$\sigma=\uparrow$ or $\downarrow$ is the spin index. We take $t_{NN}=-2.855$ eV and $t_{2NN}=-0.185$ eV for the NN and 2NN hoppings among the carbon $p_{z}$ orbitals \cite{wehling11,thomsen15}. The strength of the C-H bond is taken as $t_{CH}=5$ eV. The on-site energies of the two orbitals are taken as $\epsilon_{p}=0$ eV and $\epsilon_{s}=3.271$ eV \cite{thomsen15}. The two basis vectors are $\mathbf{a}_{1}=(\frac{1}{2},-\frac{\sqrt{3}}{2})a$ and $\mathbf{a}_{2}=(\frac{1}{2},\frac{\sqrt{3}}{2})a$, with $a\simeq2.516$ \AA \cite{sofo07}. We have also incorporated the Hubbard interaction within the carbon $p$ orbitals, and will treat it at the self-consistent mean field level (see Appendix A). This model captures the most important physics of the hydrogenation: The formation of each C-H bond passivates one carbon $p$ orbital, and the removal of each hydrogen atom releases one carbon $p$ orbital \cite{duplock04,casolo09,balog10,wehling11,huang16,lu16}.

We first study the bulk band structures of graphane. Test calculations with $U$ up to 20 eV show no sign of spin ordering or charge ordering transitions. From Fig. 1(a) for the band structures of the above model at $U=6$ eV, the hybridization between carbon $p$ orbitals and hydrogen $s$ orbitals opens a full energy gap of about 4.6 eV, the magnitude of which is within the range of existing first-principles results \cite{sofo07,sluiter03,sahin09}. Increasing $U$ to a larger value (e.g., 9.3 eV \cite{wehling11}), the band gap and the global band structure change only slightly. The considered model therefore captures correctly the transition from semimetal to insulator upon hydrogenation, the major qualitative change in the band structures from graphene to graphane.

\begin{figure}[!htb]\label{fig1}
\centering
\hspace{-5.5cm} {\textbf{(a)}}\\
\includegraphics[width=6.5cm,height=5.0cm]{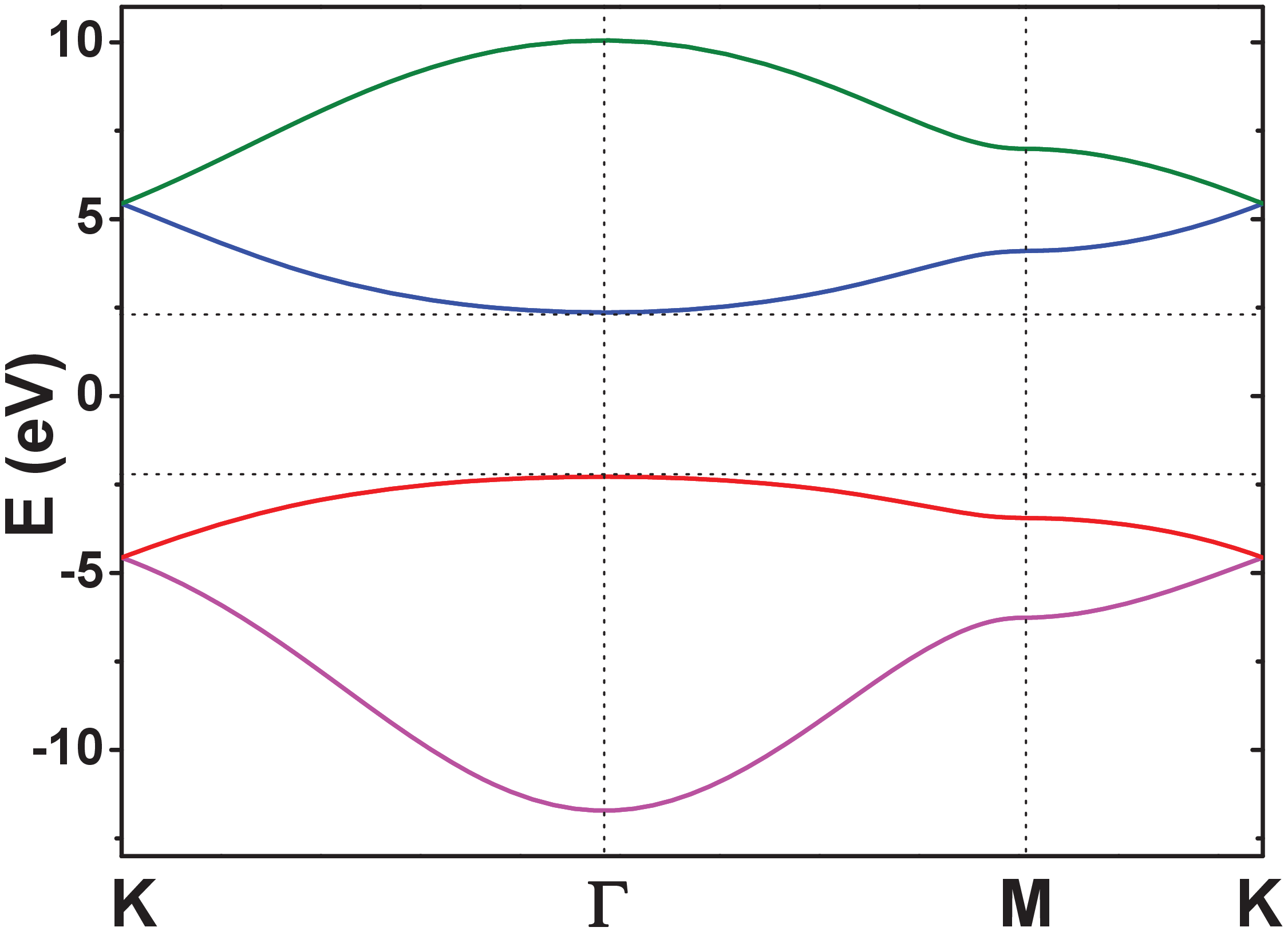} \\ \vspace{-0.05cm}
\hspace{-5.5cm} {\textbf{(b)}}\\
\includegraphics[width=6.5cm,height=5.2cm]{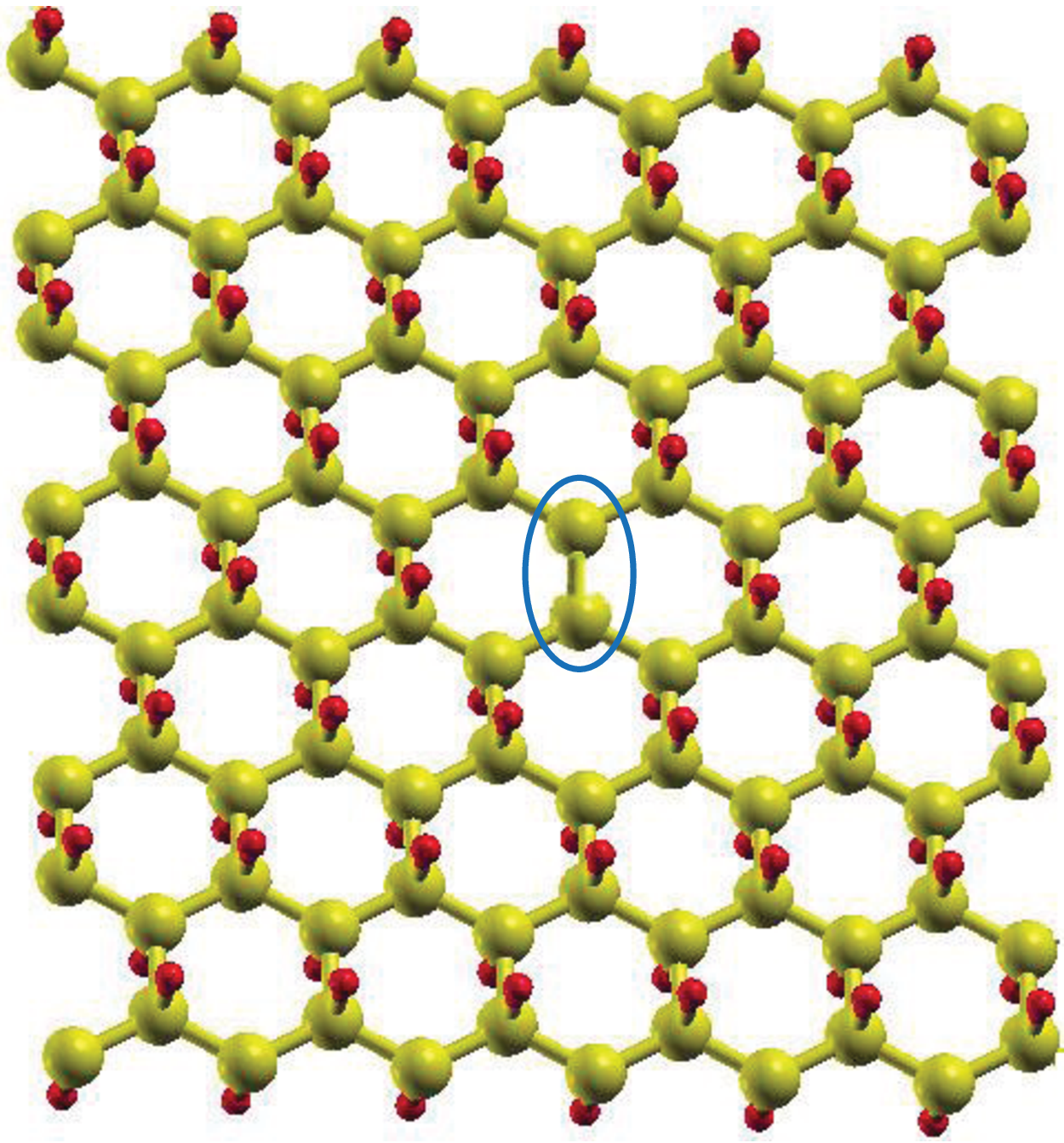}  \\
\caption{Band structures and lattice structures. (a) The band structures of the pristine graphane, for $U$=6 eV. The horizontal dotted lines label the valence band top and conduction band bottom. The vertical dotted lines label the high symmetry points in the Brillouin zone, $\boldsymbol{\Gamma}=(0,0)$, $\textbf{K}=(\frac{1}{2},\frac{\sqrt{3}}{2})\frac{4\pi}{3a}$, and $\textbf{M}=(\frac{\sqrt{3}}{2},\frac{1}{2})\frac{2\pi}{\sqrt{3}a}$. Each band is two-fold degenerate. $E$=0 eV labels the position of the chemical potential. (b) A 12$\times$8 graphane lattice with a carbon dimer defect enclosed by a blue ellipse. The larger yellow (smaller red) balls represent the carbon (hydrogen) atoms.}
\end{figure}

We next create a carbon dimer defect by removing the hydrogen atoms on two NN carbon atoms, thereby releasing two carbon $p$ orbitals. We calculate numerically the eigenstates of a graphane lattice with a single carbon dimer defect, in the real space since translational invariance is broken by the creation of the dimer. As shown in Fig.1(b), we consider a rectangular lattice with $N_{x}\times N_{y}$ C-H units (or carbon atom, if the associated hydrogen atom is removed). The $x$ ($y$) axis runs along the zigzag (armchair) direction, and $N_{x}$ ($N_{y}$) is the number of carbon sites along each zigzag chain (the number of zigzag chains). The periodic boundary condition is imposed, so that the lattice is topologically a torus and both $N_{x}$ and $N_{y}$ are even integers. The self-consistent mean field calculations (at zero temperature, unless otherwise specified) start with a uniform charge-ordered \emph{ferri}magnetic initial state, to allow the charge ordering and the (ferromagnetic or antiferromagnetic) magnetic ordering to form spontaneously (see Appendix A for more details).

\begin{figure}\label{fig2} \centering
\hspace{-2.95cm} {\textbf{(a)}} \hspace{3.8cm}{\textbf{(b)}}\\
\hspace{0cm}\includegraphics[width=4.2cm,height=3.5cm]{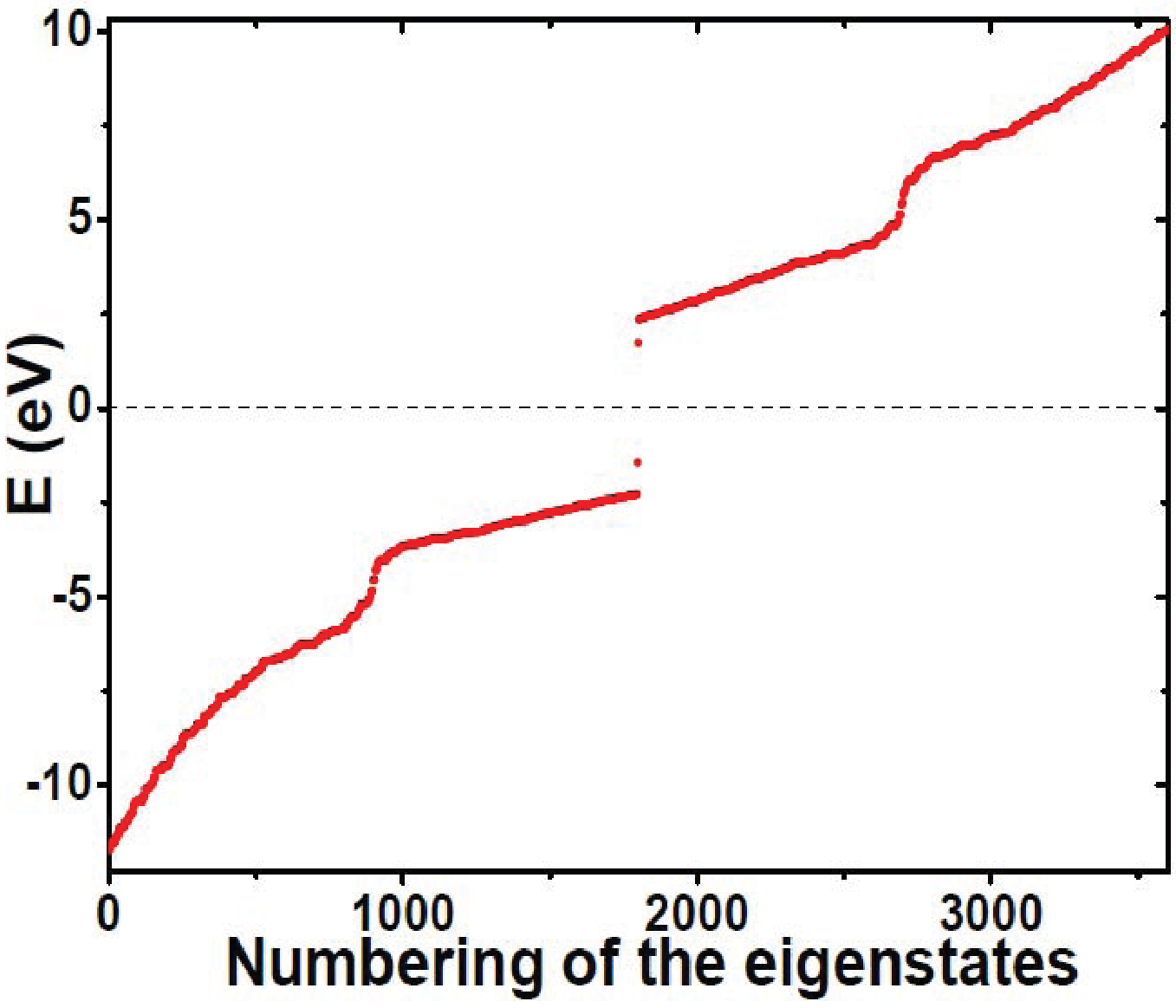}
\includegraphics[width=4.2cm,height=3.5cm]{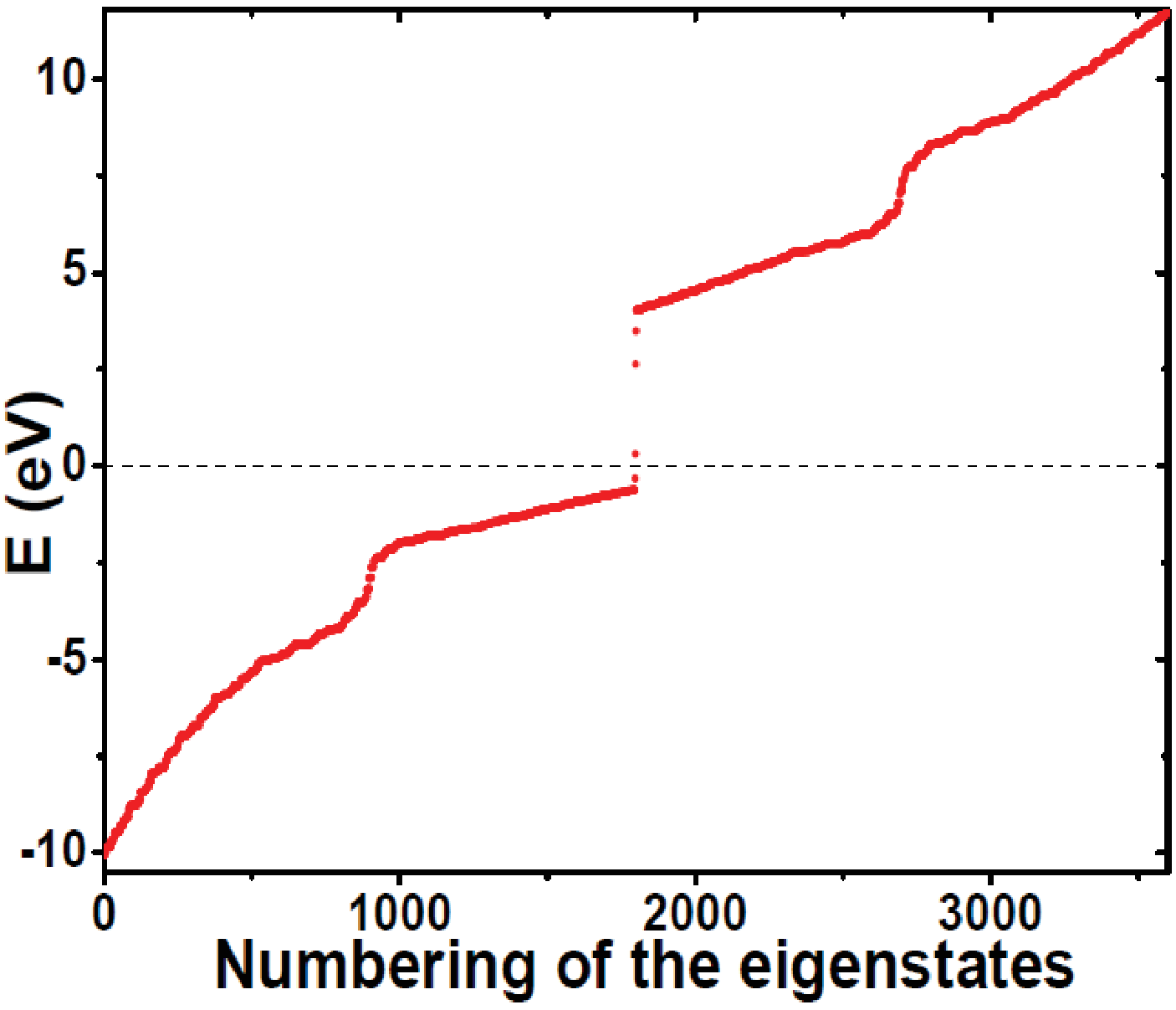}\\
\vspace{-0.10cm}
\hspace{-2.95cm} {\textbf{(c)}} \hspace{3.8cm}{\textbf{(d)}}\\
\hspace{0cm}\includegraphics[width=4.2cm,height=3.5cm]{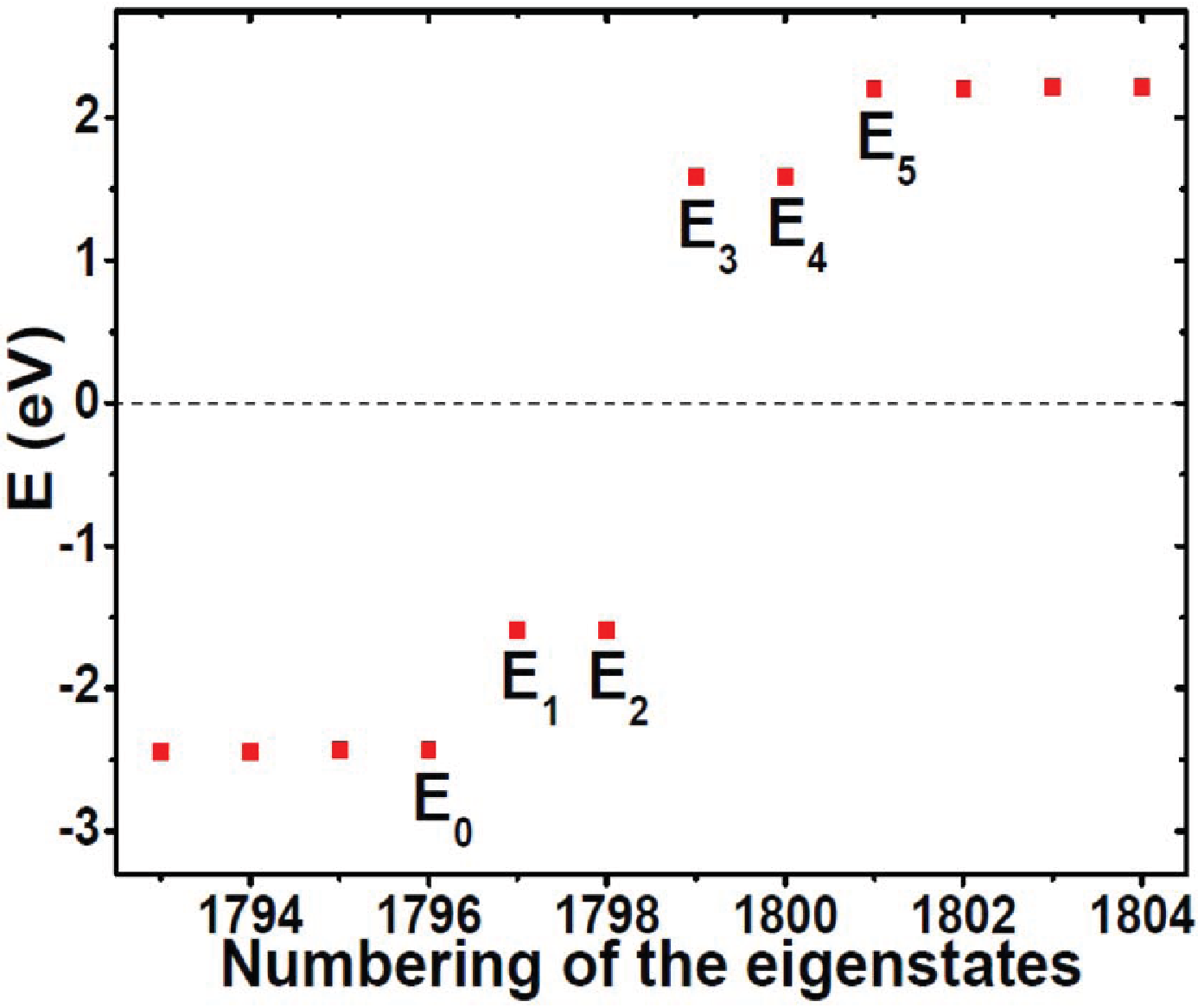}
\includegraphics[width=4.2cm,height=3.5cm]{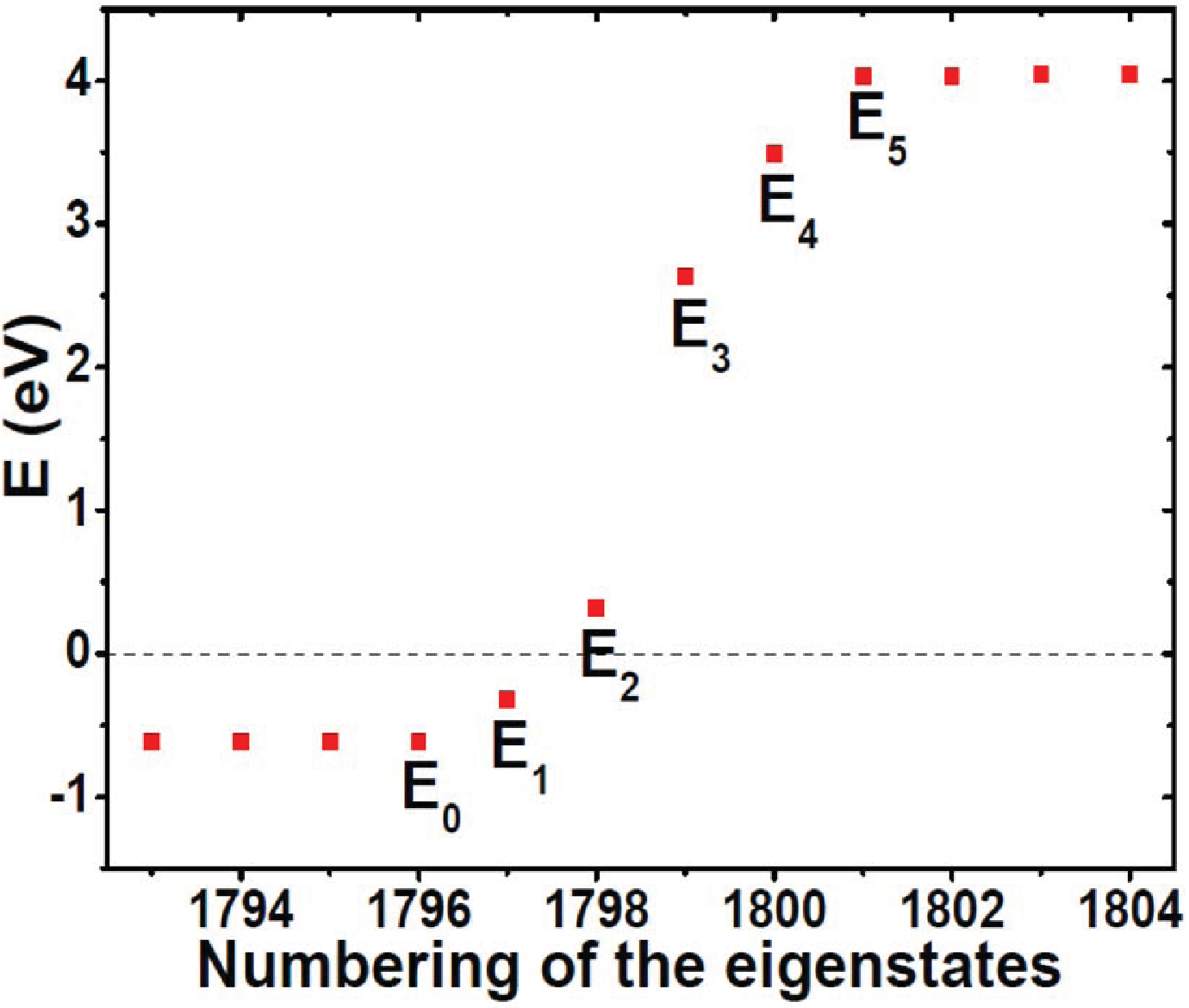}
\caption{Energy distribution of the quasiparticle states. A 30$\times$30 graphane lattice with one carbon dimer defect (two hydrogen vacancies on two NN carbon sites) is considered. $U$=6 eV. (a) and (c) are for the half-filled system. (b) and (d) are for the system with one electron less than the half-filled system. (c) and (d) are enlarged plots close to the bulk band gap. The six symbols in (c) and (d), $E_{0}$ to $E_{5}$, are the names defined for the six states right above the corresponding symbols. The horizontal dashed lines at $E$=0 eV label the positions of the chemical potential.}
\end{figure}

As shown in Figure 2 are the energy levels of a 30$\times$30 graphane lattice with a single carbon dimer defect, for $U=6$ eV. The energy level distributions are independent of the orientation of the carbon dimer bond. Besides the stoichiometric half-filled system (Figs.2a and 2c), we consider the case where one additional electron is removed from the system (Figs.2b and 2d). Additional energy levels appear inside the bulk band gap for both cases. For the half-filled system, there are two in-gap energy levels, with each one two-fold degenerate. The four in-gap defect states are posited right at the center of the full sequence of all the states. That is, among the 30$\times$30$\times$(2+2)-2$\times$2=3596 states of the 30$\times$30 lattice with two missing hydrogen atoms, the in-gap states are numbered as the 1797th state to the 1800th state. To be clear, we have defined on Figs. 2(c) and 2(d) the valence band top as $E_{0}$, the four defect states as $E_{1}$ to $E_{4}$ in an order of increasing energy, and the conduction band bottom as $E_{5}$. So, in the ground state of the half-filled system, we have one occupied and one empty two-fold degenerate in-gap energy levels (Fig. 2c). By removing one electron from the half-filled system, the two-fold degeneracies in the in-gap energy levels are broken and we now have four nondegenerate in-gap energy levels (Fig. 2d). The defect states are now fully spin polarized and the ground state has a magnetic moment from the occupied lowest-energy in-gap defect state ($E_{1}$). The wave functions of the in-gap states center around the two carbon sites of the dimer. Away from the 2NN sites of the carbon dimer sites, the weights of the in-gap states decay rapidly to be negligible.

\begin{figure}\label{fig3} \centering
\includegraphics[width=7.5cm,height=5.0cm]{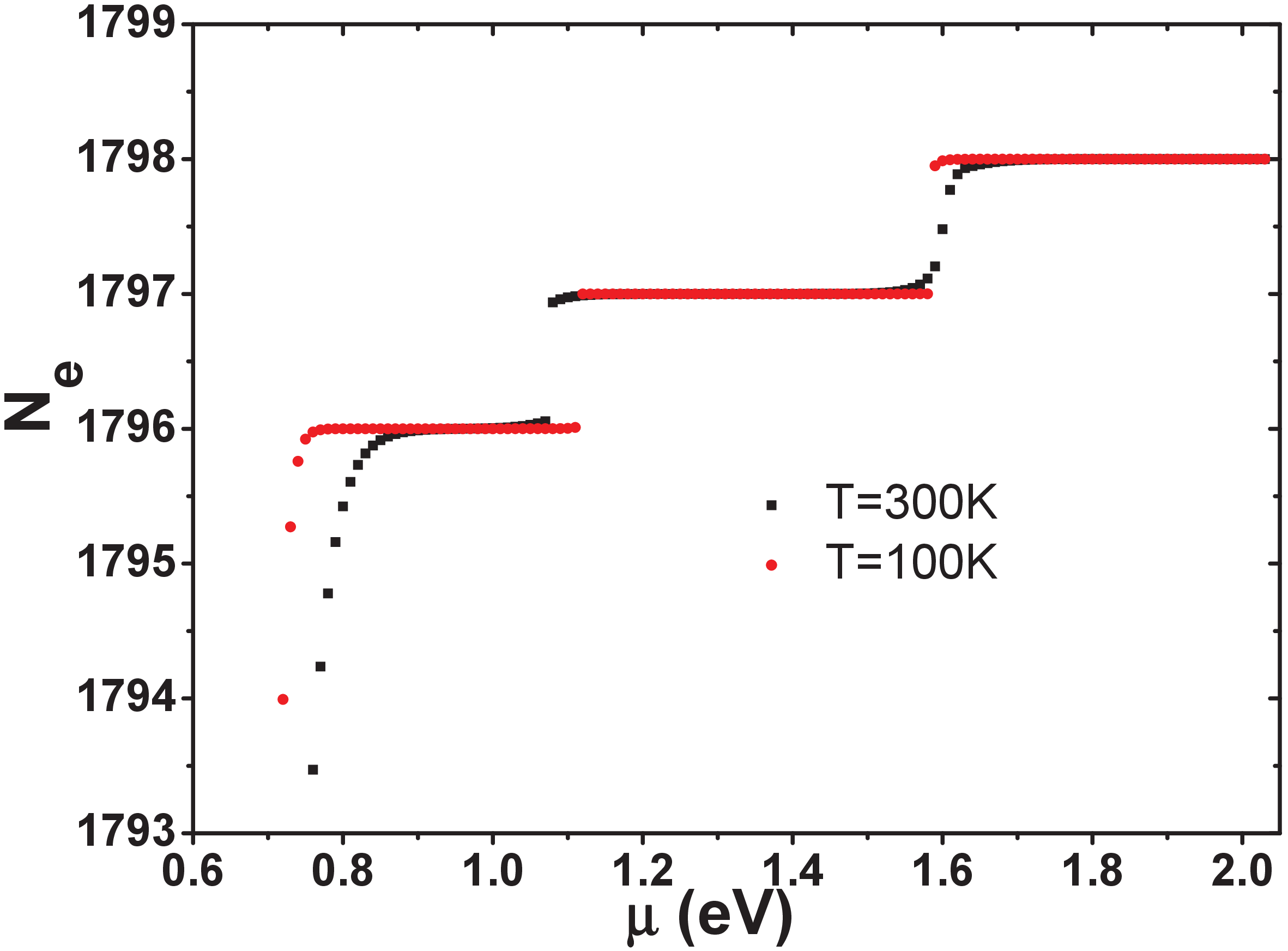}
\caption{Total number of electrons ($N_{e}$) as a function of the chemical potential ($\mu$), in a 30$\times$30 graphane lattice with one carbon dimer defect. $U$=6 eV. The self-consistent mean field calculations (at 300 K and 100 K) are performed in the presence of a static magnetic field $B_{0}$, which gives a Zeeman energy of $\mu_{0}B_{0}=0.001$ eV, with $\mu_{0}$ the Bohr magneton.}
\end{figure}

The case with one electron removed from the system per carbon dimer (i.e., Figs.2b and 2d) can be attained by fine tuning a gate voltage in the presence of a static magnetic field $B_{0}$, or by grabbing an electron from the carbon dimer with a positively charged scanning tunneling probe. Here, we illustrate the first mechanism in Figure 3, at both the room temperature (300 K) and a lower temperature (100 K). A uniform static magnetic field is applied to give a Zeeman energy of $\mu_{0}B_{0}=0.001$ eV, where $\mu_{0}$ is the Bohr magneton. The magnetic field is assumed to be oriented along the $z$ axis, perpendicular to the graphane lattice. The magnetic field on one hand splits the two-fold degeneracy of the in-gap states in the half-filled system, on the other hand fixes the spin quantization axis along the $z$ direction. The relevant fine tuning of the electron occupation number is associated with the jump of $N_{e}$ from $1798$ to $1797$. Importantly, the width of the chemical potential range for which the total electron number is $1797$ is determined not by the Zeeman energy, but by the energy separation between $E_{1}$ and $E_{2}$ in the case with one electron less than stoichiometry.

To test the robustness of the in-gap defect states, $E_{1}$ to $E_{4}$, we study their dependence on $U$ and the lattice size. The dependence on $U$ of several characteristic energy intervals are shown in Fig. 4. First of all, for moderate and small correlation strength $U<9$ eV ($U<10$ eV), the half-filled (one electron less than half-filled) system always has four well-defined in-gap defect states. While we have $E_{2}-E_{1}=E_{4}-E_{3}=0$ for the half-filled system (Fig. 4a), removing one electron from the system breaks this degeneracy for all $U>0$ (Fig. 4b). In the system with one electron less than half filling and for $U<10$ eV, $E_{1}$ and $E_{3}$ are found to have the same spin, which is opposite to the spin of $E_{2}$ and $E_{4}$. We have also checked for $U=6$ eV that, as the size of the graphane lattice becomes larger than 12$\times$8 (up to 64$\times$64), there are only very tiny changes in the in-gap energy levels and their positions with respect to the bulk conduction and valence bands. Therefore, the in-gap defect states for both the half-filled system and the one electron less than half-filled system are robust qualitative features of the graphane lattice with a carbon dimer defect.

We now construct the low-energy effective model of the defect states, for $U\le9$ eV. Because the graphane lattice under periodic boundary conditions has the inversion symmetry both before and after the creation of a carbon dimer defect (see Fig.1b), the defect states should have definite parities. Denoting the eigen-ket of the defect level $E_{i}$ as $|i\rangle$ ($i=1,...,4$) and constructing the parity operator $\hat{P}$ with respect to the center of the carbon dimer bond, we find through numerical calculations that the two lower ($E_{1}$ and $E_{2}$) and two upper ($E_{3}$ and $E_{4}$) defect states have separately even and odd parities. That is, $\hat{P}|i\rangle=|i\rangle$ for $i=1$ and 2, and $\hat{P}|i\rangle=-|i\rangle$ for $i=3$ and 4. In going from the half-filled system to the system with one electron less than half filling, the two-fold spin degeneracy in the two in-gap energy levels are broken, but the parities of the states do not change. Denoting the creation operators for the states with even (odd) parity and $\sigma$ ($=\uparrow$, $\downarrow$) spin as $a^{\dagger}_{\sigma}$ ($b^{\dagger}_{\sigma}$), namely $a^{\dagger}_{\uparrow}$ ($a^{\dagger}_{\downarrow}$) creates the $E_{1}$ ($E_{2}$) state and $b^{\dagger}_{\uparrow}$ ($b^{\dagger}_{\downarrow}$) creates the $E_{3}$ ($E_{4}$) state, we write the low-energy effective model $\hat{H}_{C2}$ for the in-gap defect states as
\begin{equation}
\frac{\Delta}{2}\sum\limits_{\sigma}(b^{\dagger}_{\sigma}b_{\sigma}-a^{\dagger}_{\sigma}a_{\sigma})
+\frac{m}{2}(a^{\dagger}_{\downarrow}a_{\downarrow}-a^{\dagger}_{\uparrow}a_{\uparrow})
+\frac{m'}{2}(b^{\dagger}_{\downarrow}b_{\downarrow}-b^{\dagger}_{\uparrow}b_{\uparrow}).
\end{equation}
$\Delta=(E_{3}+E_{4})/2-(E_{1}+E_{2})/2$. $m$ and $m'$ are respectively the spin splitting energies of the two even-parity and two odd-parity states. For the half-filled system, we have $m=m'=0$. For the system with one electron less than half filling, we have $m>0$ and $m'>0$. By applying a static magnetic field along the $z$ axis, during the process of removing one electron from the carbon dimer, we can set the spin quantization axis perpendicular to the graphane plane. Introduce the basis operator for the subspace expanded by the four defect states as $\phi^{\dagger}=[a^{\dagger}_{\uparrow},a^{\dagger}_{\downarrow},
b^{\dagger}_{\uparrow},b^{\dagger}_{\downarrow}]$. Define $\sigma_{\alpha}$ and $s_{\alpha}$ ($\alpha=x,y,z$) as the Pauli matrices in the subspace of parity and spin, respectively. In this new basis, we write $\hat{H}_{C2}=\phi^{\dagger}H_{C2}\phi$, with
\begin{equation}
H_{C2}=-\frac{\Delta}{2}\sigma_{z}\otimes s_{0}-\frac{m+m'}{4}\sigma_{0}\otimes s_{z}
-\frac{m-m'}{4}\sigma_{z}\otimes s_{z},
\end{equation}
where $\sigma_{0}$ and $s_{0}$ are separately unit matrices in the parity and spin subspaces.

\begin{figure}\label{fig4} \centering
\hspace{-2.95cm} {\textbf{(a)}} \hspace{3.8cm}{\textbf{(b)}}\\
\hspace{0cm}\includegraphics[width=4.2cm,height=3.5cm]{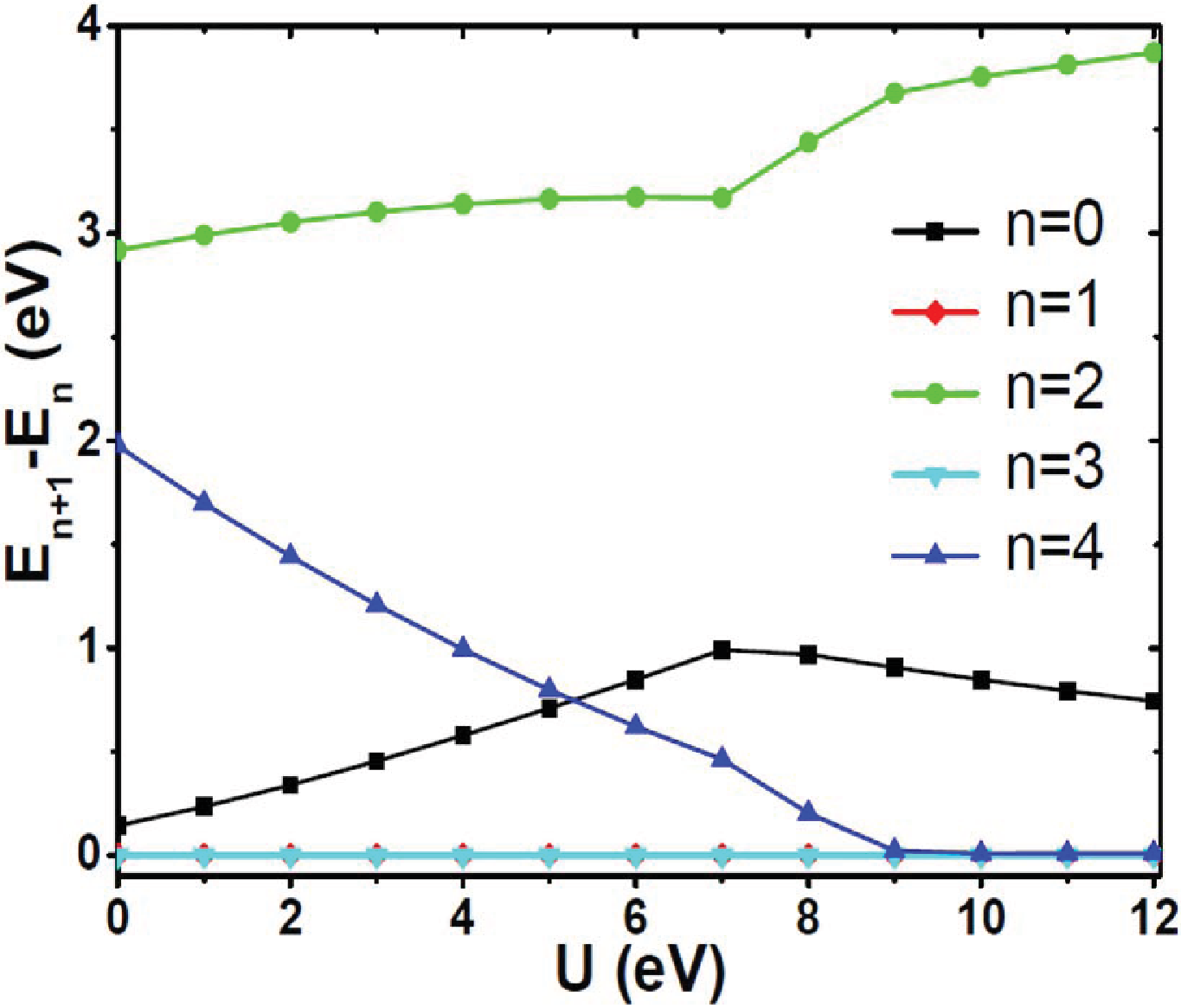}
\includegraphics[width=4.2cm,height=3.5cm]{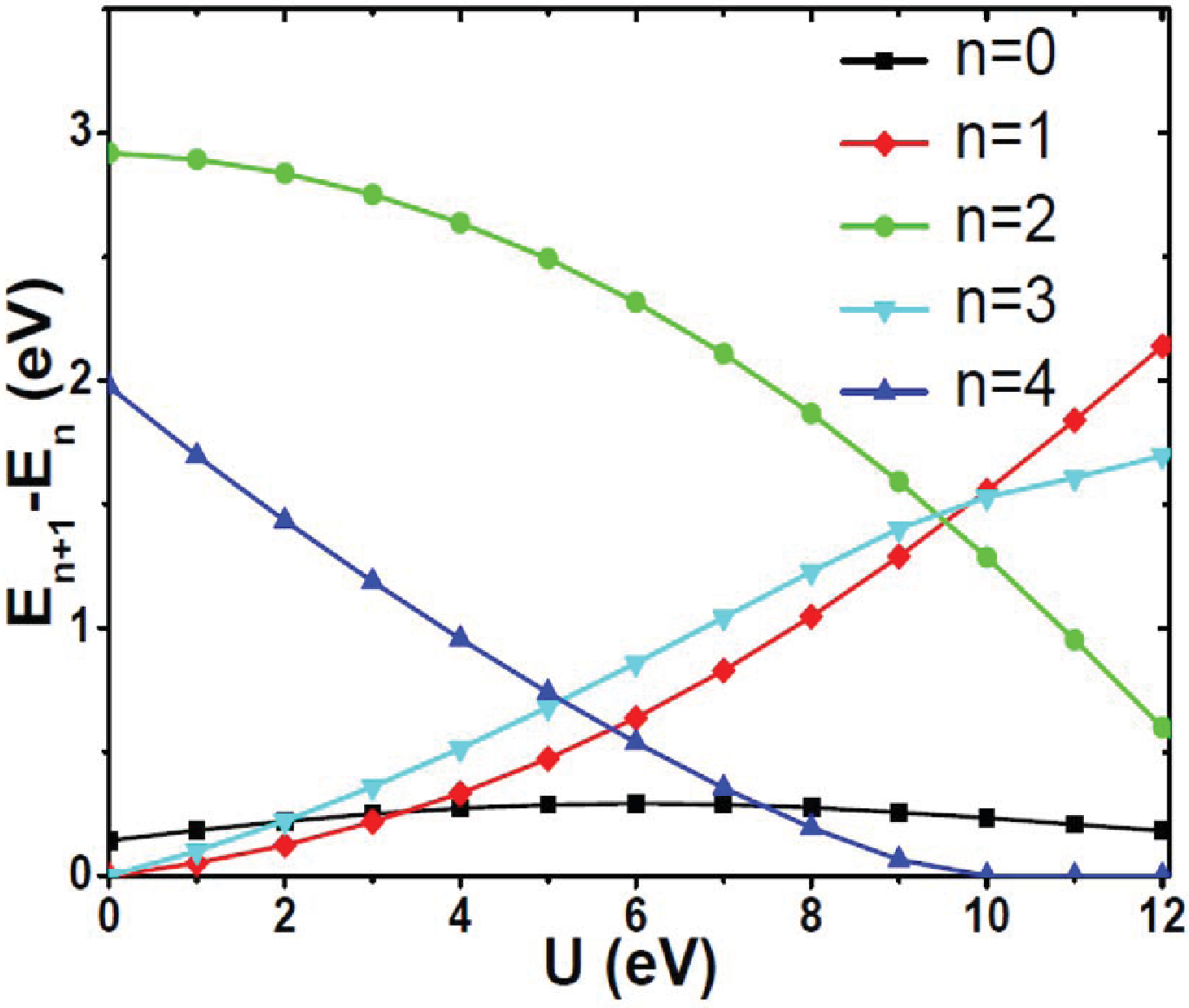}\\
\vspace{-0.10cm}
\hspace{-2.95cm} {\textbf{(c)}} \hspace{3.8cm}{\textbf{(d)}}\\
\hspace{0cm}\includegraphics[width=4.2cm,height=3.5cm]{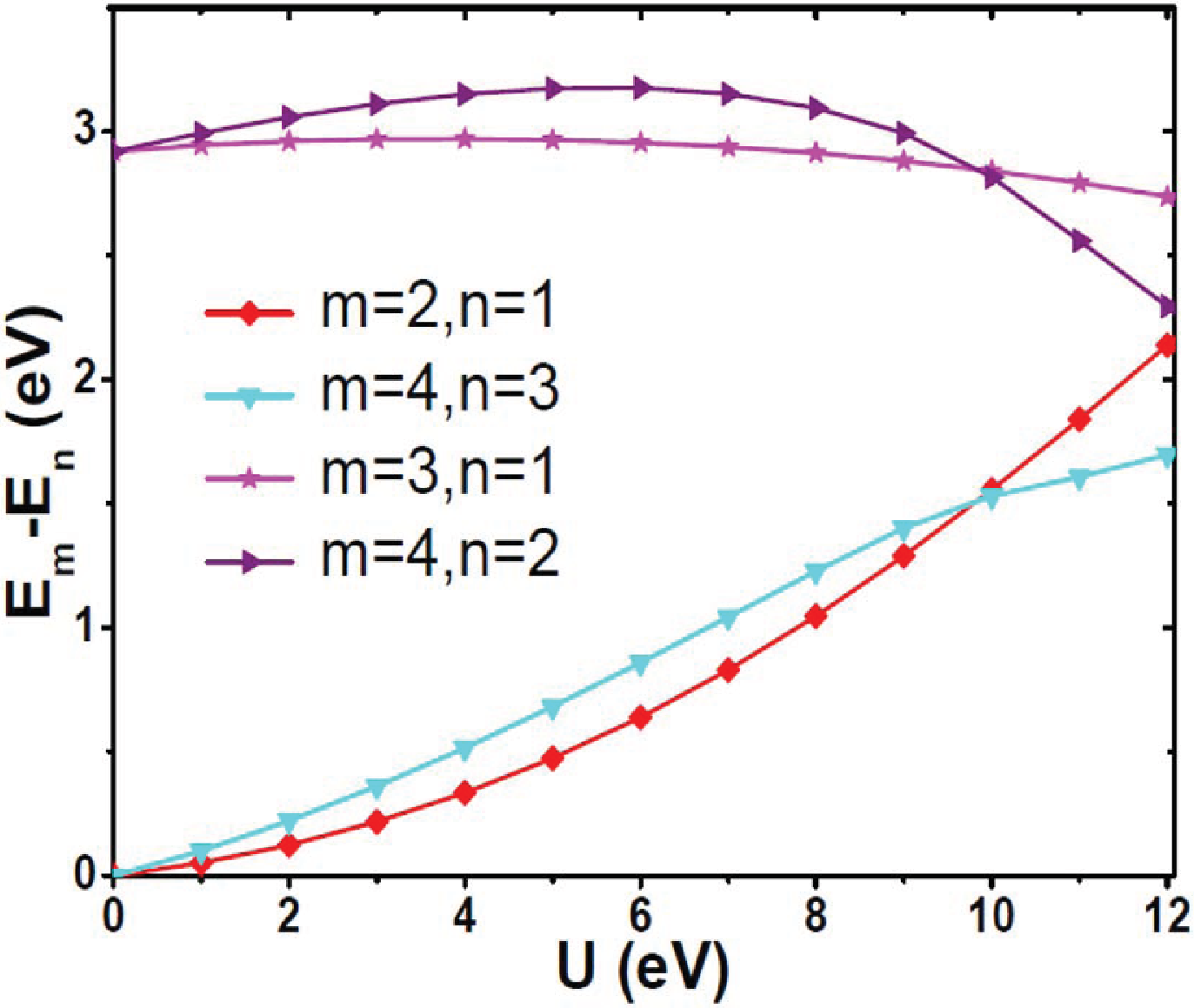}
\includegraphics[width=4.2cm,height=3.5cm]{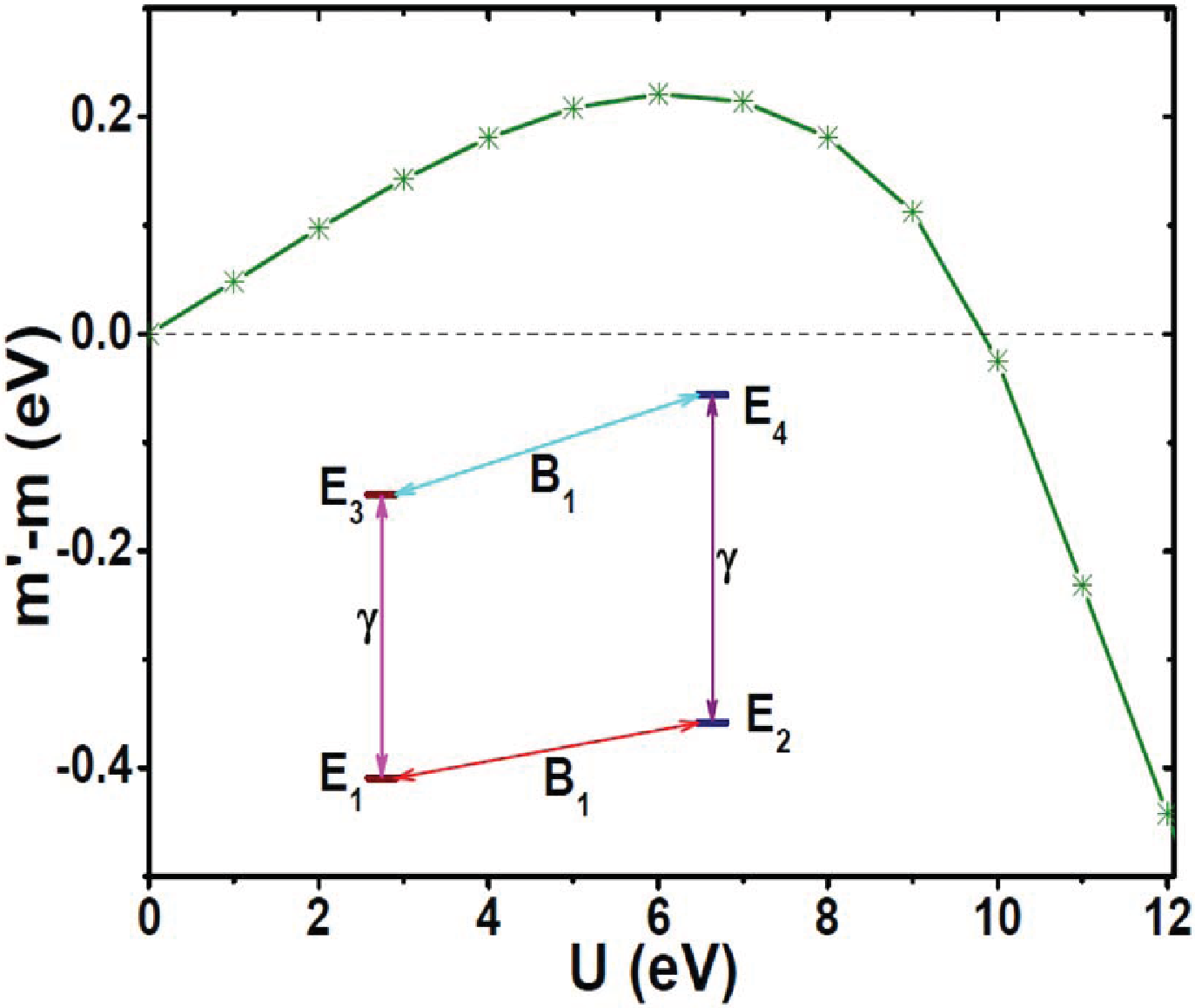}
\caption{Dependence of the in-gap defect states on the Hubbard interaction U. A single carbon dimer defect is created inside a 30$\times$30 graphane lattice. The subindices of the energies follow the definitions of Figs. 2c and 2d. (a) is for the half-filled system. The curves for $n+1$=2 and $n+1$=4 coincide, which are zero for all $U$. (b), (c), and (d) are for the system with one electron less than the half-filled system. The inset of (d) is the schematic energy level structure of the defect states, for $U\le9$ eV. It also shows the four possible transitions, corresponding to the four energy intervals on Fig. 4(c), among the four defect states. $\gamma$ and $B_{1}$ represent the optical field and the \emph{ac} magnetic field, respectively.}
\end{figure}

\section{\label{manipulations}Quantum manipulations of defect states}

Transitions between opposite-parity (opposite-spin) states of the same spin (parity) could be triggered by an optical field (\emph{ac} magnetic field) through the electric dipole (magnetic dipole) transition. These manipulations are summarized as the following driving Hamiltonian \cite{sakuraibook}
\begin{eqnarray}
\hat{H}_{d}&=&-\mu_{0}B_{0}\sum\limits_{\sigma}\sigma(a^{\dagger}_{\sigma}a_{\sigma}+b^{\dagger}_{\sigma}b_{\sigma}) \notag \\
&&-\mu_{0}B_{1}\cos(\omega_{1}t)(a^{\dagger}_{\uparrow}a_{\downarrow}+b^{\dagger}_{\uparrow}b_{\downarrow}+\text{H.c.}) \notag \\
&&+2i\gamma\cos(\omega t)\sum\limits_{\sigma}(a^{\dagger}_{\sigma}b_{\sigma}-b^{\dagger}_{\sigma}a_{\sigma}).
\end{eqnarray}
In the basis $\phi^{\dagger}$, we have $\hat{H}_{d}=\phi^{\dagger}H_{d}\phi$ where 
\begin{equation}
H_{d}=-\mu_{0}B_{0}\sigma_{0}\otimes s_{z}-\mu_{0}B_{1}\cos(\omega_{1}t)\sigma_{0}\otimes s_{x}
-2\gamma\cos(\omega t)\sigma_{y}\otimes s_{0}.
\end{equation}
$B_{0}$, $B_{1}$, and $\gamma$ are implicitly time-dependent, nonzero in certain time intervals. $B_{0}$ is the \emph{dc} magnetic field along the $z$ direction that fixes the spin quantization axis of the system with one electron less than half filling. $B_{1}$ is the magnitude of an \emph{ac} magnetic field of frequency $\omega_{1}$ and directed along the $x$ direction. $\gamma$, which denotes the strength of the electric dipole transition, depends on the polarization vector and the intensity of the laser pulse. Both the $B_{1}$ term and the $\gamma$ term induce two Rabi oscillations (see the inset of Fig. 4d). For the system with one electron less than half filling, the two transitions induced by the same stimulus have different resonant frequencies. For $B_{1}$, the two resonant frequencies are $m$ and $m'$. For $\gamma$, the two resonant frequencies are $\Delta+(m-m')/2$ and $\Delta+(m'-m)/2$. As is shown in Fig.4(d), $m\neq m'$ holds for general $U$.

Analysis of the quantum manipulations could be simplified by introducing the rotating frame \cite{rabi54}. Consider the rotating frame defined by (in the basis $\phi^{\dagger}$, we assume $\hbar=1$ hereafter)
\begin{equation}
H_{0}=-\frac{\omega}{2}\sigma_{z}\otimes s_{0}-\frac{\omega_{1}}{2}\sigma_{0}\otimes s_{z}.
\end{equation}
The effective rotating frame Hamiltonian for $H=H_{C2}+H_{d}$ with respect to $H_{0}$ is \cite{nielsenbook}
\begin{eqnarray}
\tilde{H}&=&e^{iH_{0}t}He^{-iH_{0}t}-H_{0}=-\frac{m-m'}{4}\sigma_{z}\otimes s_{z}   \notag  \\
&&-(\frac{m+m'}{4}+\mu_{0}B_{0}-\frac{\omega_{1}}{2})\sigma_{0}\otimes s_{z}-\frac{\mu_{0}B_{1}}{2}\sigma_{0}\otimes s_{x}  \notag \\
&&-\frac{\Delta-\omega}{2}\sigma_{z}\otimes s_{0}-\gamma\sigma_{y}\otimes s_{0}.
\end{eqnarray}
The evolution of the defect states is thus driven by $\tilde{H}$ and $H_{0}$, which are both simpler than $H$. The terms in the second line and third line of Eq.(7) control separately the evolutions in the subspace of the two spins and the two parities. The first term of $\tilde{H}$ couples the evolutions in the two channels. In arriving at Eq.(7), we have applied the rotating-wave approximation to discard the high-frequency terms \cite{sakuraibook,rabi54}. Because $m'\neq m$, we can realize approximately the four transitions illustrated in Fig. 4(d) one at a time, by tuning the frequencies of the laser field and (or) the \emph{ac} magnetic field, and by controlling the lengths of the pulse sequences. For example, by tuning $\omega_{1}=m'$ and turning off the optical field, we can realize a quantum operation that exchanges $E_{3}$ and $E_{4}$.


A system can realize universal quantum computation under the premise that arbitrary one-qubit unitary gates (i.e., unitary quantum operations) and at least one nontrivial two-qubit gate (e.g., the controlled-NOT gate) can be implemented with it \cite{barenco95,deutsch95,lloyd95,bremner02,nielsenbook}. We explain in what follows why the graphane with carbon dimer defects meets this requirement. More details are contained in Appendix B.

First consider the one qubit unitary gates. For the half-filled system, we ignore the spin degree of freedom and the defect states constitute a parity qubit. Tuning the defect states with the optical field, the model consists of the last two terms of Eq.(7). For $\omega=\Delta$, the Rabi oscillation induced by the $\gamma$ term act as a $Y$-rotation gate. Together with the $Z$-rotation gate associated with the free evolution driven by $H_{C2}(m=m'=0)$, we can achieve arbitrary unitary transformations on the parity qubit \cite{nielsenbook}.

For the system with one electron less than half filling (per carbon dimer), let us focus on the two low-energy even-parity states, which constitute a single spin qubit. Now, the $Z$-rotation gate is realized by the $m$ term of $H_{C2}$, and the \emph{ac} magnetic field $B_{1}$ with frequency $\omega_{1}=m+2\mu_{0}B_{0}$ actualizes the $X$-rotation gate for this spin qubit. Again, we can combine these two kinds of operations to perform arbitrary single-qubit unitary gates on this spin qubit \cite{nielsenbook}.

We consider next the two-qubit gates. A two-qubit gate is encoded in any four states that can be manipulated in a nontrivial manner at the two-qubit level \cite{nielsenbook,kessel99,kessel02,kiktenko15}. In the present defective graphane lattice, the four states may be the four nondegenerate defect states of a carbon dimer with one electron less than half filling, or the four doubly degenerate defect states of two (stoichiometric) half-filled carbon dimers. For the former case, the four transitions shown in Fig.4(d) have different resonant frequencies and can have separate manipulations. The controlled-NOT gate, which performs a swapping of two (e.g., $E_{3}$ and $E_{4}$) among the four in-gap states, is certainly realizable (see Appendix B for more details). For the latter case, we have to couple the states of the two carbon dimers. Compared to local couplings mediated by the lumped electronic circuits or fine tuning the inter-dimer distances, the distributed quantum computing, which entangles spatially separated qubits (i.e., the half filled carbon dimers) via linear optics \cite{barrett05,knill01,duan01,majer07,gorman16}, seems more flexible and promising. The application of the distributed quantum computing scheme \cite{barrett05,knill01} to achieve nontrivial two-qubit gates in the system with half-filled carbon dimers will not be explained here.

\section{\label{framework}A scalable 2D quantum computing platform}

Based on the above analyses, graphane with carbon dimer defects appear to be a unique 2D platform for quantum computer, similar to the silicon wafer for digital computer. Suppose we have prepared an ideal sample of graphane, we then proceed by making analogy between this graphane sheet and the silicon wafer of the semiconductor industry. A carbon dimer is an atomic-size qubit (or, qudit, the multilevel generalization of qubit \cite{kessel99,kessel02,kiktenko15}) immersed within the insulating bulk matrix of graphane, analogous to transistors in the silicon wafer. By creating more carbon dimers, controlling the distances and relative dimer bond orientations between different dimers, we get a qubit network whose property is controlled by design. The graphane lattice with properly created carbon dimer defects then serves as an ideal scalable quantum computing network. We discuss in what follows several relevant issues.

For the preparation of a fiducial initial state \cite{nielsenbook}, we may take the ground state or a state connected to the ground state by a Rabi flopping as the initial state. Since the energy separation between the occupied and the empty defect states has a magnitude of $0.5$ eV ($3$ eV) for the system with one electron less than half filling (half-filled system), the mean-field parameters and the energy level structures at zero temperature are barely changed at a temperature about $300$ K, which is confirmed by explicit numerical calculations. We therefore expect to achieve high-fidelity quantum control at room temperature, because the thermal fluctuation energy is at least one order of magnitude smaller than the above energy separation.

The decoherence may be alleviated by suspending the graphane lattice to isolate it from the environment, and by sweeping the gates at a speed faster than the decoherence processes. The suspended graphane lattice follows directly from the suspended graphene lattice, which is the natural motherboard to produce the graphane. For resonant Rabi transitions, the sweeping speed is determined by $\gamma$ or $\mu_{0}B_{1}$ \cite{rabi54,nielsenbook,khaneja01} (see Appendix B). By increasing the intensity of the laser pulse or the strength of the \emph{ac} magnetic field, it is possible to realize gates fast compared to the relevant decoherence processes.

The spatial locality of the in-gap states implies that we can process a tiny graphane single crystal into a chip with an extremely high areal density of carbon dimer defects. Suppose we create one carbon dimer in every $12\times8$ supercell (e.g., according to Fig.1b), the neighboring carbon dimer defects are electronically isolated according to our numerical calculations for $U=6$ eV. Then, a nanometer-size graphane flake may host a single carbon dimer, and a graphane flake of an area about 1$\mu$m$\times$1$\mu$m can contain more than $720$$\times$$650$ isolated carbon dimer defects. This estimation is however over-optimistic, because it is hard to reduce the cross section of a beam of electromagnetic wave much smaller than its wavelength. In the semiconductor industry, the immersion lithography was introduced to reduce the feature size (e.g., 22 nm) of the integrated circuit well below the wavelength of the light (i.e., 193 nm) used in the lithography \cite{lin02}. In normal conditions, to independently manipulate different carbon dimers, the separation between NN carbon dimers should be larger than the wavelength of the stimuli. For $U\simeq6$ eV, the resonant transitions shown in Fig.4(d) correspond to \emph{ac} magnetic fields of about 2.5 $\mu$m in wavelength (i.e., infrared light) and optical fields of about 0.4 $\mu$m in wavelength (i.e., violet visible light). Therefore, a graphane sample of an area about 1mm$\times$1mm is able to host at least $100$$\times$$100$ well-separated carbon dimers. This might already be enough to illustrate the large-scale quantum computations \cite{preskill98}. If a large computational overhead turns out to be necessary to perform the quantum error corrections \cite{fowler12}, a centimeter-size graphane sample should contain sufficient number of carbon dimers for large-scale quantum computation.

Finally, we propose methods of creating the carbon dimers in graphane. One approach is to remove the hydrogen atoms mechanically or electrically by an atomic-size scanning tunneling probe \cite{lyding94,randall09,pavlicek17,eigler90,shen95}. In application, we flip the graphane lattice after removing the target hydrogen atoms on one side of the sample, and then remove the target hydrogen atoms on the opposite side of the sample. A related technique, lithography with the electron beam from a scanning electron microscopy \cite{lee15}, might also be useful if the electron beam has a diameter in the angstrom scale. Another prospective method is the C-H activation, a technique of breaking and functionalizing the C-H bonds in organic molecules \cite{liao16,zhang16,pd1,pd2}. One implementation of this method is to remove hydrogen atoms by forming carbon-metal bonds and then remove the metallic atoms. This method can be combined with the first method, by preparing the scanning tunneling probe with the metal that can activate the C-H bonds (e.g., Pd \cite{pd1,pd2}). If the tip of the probe is stiff enough (e.g., by alloying), we may remove directly the metallic tip without leaving a metal atom above the carbon atom. To make the first and the last methods feasible, the tip of the scanning tunneling probe must be atomically sharp, to ensure that only the hydrogen atom underneath the probe is removed. This requirement, although seemingly demanding, is realistic because tailoring microstructures at the single-atom level with scanning tunneling microscope have been achieved long ago \cite{lyding94,randall09,pavlicek17,eigler90,shen95}. To increase the accuracy of real-space positioning associated with flipping the graphane lattice, we may embed the graphane sample into a rigid frame, which can be rotated around a fixed axis in a finely controlled manner. In addition, besides controlling the accuracy of the flipping process, we may run a (partial) scanning tunneling microscopy of the graphane sample after the flipping. This will allow us to map the distribution of the H vacancies introduced to the first side of the graphane lattice. Based on this knowledge, we may proceed to remove accurately the target H atoms on the second side of the graphane sample. Finally, instead of flipping the graphane sample, we may place the graphane lattice and the frame holding it vertically, and then remove the H atoms on the two sides of the graphane sample simultaneously with two individual probes, in terms of a two-probe version of the scanning tunneling microscopy \cite{kolmer17,voigtlander18}.

\section{\label{conclude}summary}

In summary, we have found robust in-gap states in graphane with carbon dimer defects. Each carbon dimer defect introduces four defect states, which fall into two doubly degenerate energy levels in the stoichiometric half-filled system and are all nondegenerate in the system where one additional electron is removed from each carbon dimer. By encoding the single-qubit and two-qubit systems in these in-gap states, we can realize universal quantum computation at ambient conditions. Being intrinsic defects of a crystalline solid-state material, different carbon dimers have identical energy structures and can realize exactly the same quantum gate when acted on by the same operation. The spatial locality of the in-gap states ensures the scalability of the carbon dimers in graphane. These features testify graphane as a unique two-dimensional platform to realize future large-scale quantum computations.

\begin{acknowledgments}
This work was supported by the NSFC Grants No.11147108 and No.11574108, and the Robert A. Welch Foundation under Grant No. E-1146. Part of the calculations were performed at the Center for Advanced Computing and Data Systems at the University of Houston.
\end{acknowledgments}\index{}

\begin{appendix}

\section{The self-consistent mean field calculations}

In Eq. (1) of the main text, the only interaction term is the on-site Hubbard interaction within the $p$-orbitals of the carbon atoms. The mean-field study of the model begins with the following decoupling to the Hubbard term
\begin{eqnarray}
&&U\sum\limits_{i}c^{\dagger}_{i\uparrow}c_{i\uparrow}c^{\dagger}_{i\downarrow}c_{i\downarrow}\simeq   \\
&& U\sum\limits_{i}[\langle c^{\dagger}_{i\uparrow}c_{i\uparrow}\rangle c^{\dagger}_{i\downarrow}c_{i\downarrow}
+c^{\dagger}_{i\uparrow}c_{i\uparrow}\langle c^{\dagger}_{i\downarrow}c_{i\downarrow}\rangle
-\langle c^{\dagger}_{i\uparrow}c_{i\uparrow}\rangle \langle c^{\dagger}_{i\downarrow}c_{i\downarrow}\rangle].  \notag
\end{eqnarray}
$\langle\hat{A}\rangle$ is the expectation value of the $\hat{A}$ operator. At zero temperature, $\langle \hat{A}\rangle=\langle GS|\hat{A}|GS\rangle$ is defined in terms of the ground state $|GS\rangle$. At nonzero temperature $T>0$ K, $\langle \hat{A}\rangle=\text{Tr}\{e^{-\beta \hat{K}}\hat{A}\}/Z$, where $Z=\text{Tr}\{e^{-\beta \hat{K}}\}$ and $\beta=1/(k_{B}T)$ ($k_{B}$ is the Boltzmann constant). $\hat{K}=\hat{H}-\mu\hat{N}$ and $\mu$ is the chemical potential. $\hat{N}=\sum_{i}(c^{\dagger}_{i\uparrow}c_{i\uparrow}+c^{\dagger}_{i\downarrow}c_{i\downarrow} +\eta_{i}h^{\dagger}_{i\uparrow}h_{i\uparrow}+\eta_{i}h^{\dagger}_{i\downarrow}h_{i\downarrow})$ is the operator for the total number of electrons, where $\eta_{i}=0$ for the hydrogen vacancy sites and $\eta_{i}=1$ otherwise. $\text{Tr}\{\cdots\}=\sum_{n} \langle n|\cdots|n\rangle$ is the trace operation over the complete set of eigenstates. For simplicity in notation, we define the two mean-field parameters introduced for the $i$-th carbon site as $n_{i\uparrow}=\langle c^{\dagger}_{i\uparrow}c_{i\uparrow}\rangle$ and $n_{i\downarrow}=\langle c^{\dagger}_{i\downarrow}c_{i\downarrow}\rangle$. By substituting Eq. (A1) into Eq. (1) of the main text, we obtain the mean-field Hamiltonian. From the above definitions, the mean-field Hamiltonian and its eigenspectrum depends on the mean-field parameters $n_{i\sigma}$ ($\sigma=\uparrow,\downarrow$), the mean-field parameters are defined by the eigenstates of the mean-field Hamiltonian. This mutual dependence defines a self-consistency loop for the mean-field parameters and the mean-field Hamiltonian.

The self-consistent calculation begins by assigning an initial value to each of the independent mean-field parameters, $n_{i\sigma}=n^{(0)}_{i\sigma}$ ($\sigma=\uparrow, \downarrow$). Substituting these mean-field parameters to Eq. (A1) and replacing the Hubbard term in Eq. (1) by Eq. (A1), we get the mean-field Hamiltonian. This mean-field model is then diagonalized, which gives its full eigenspectrum and the corresponding eigenvectors. Then we determine $\mu$ by requiring the correct total number of electrons in the system, $\langle\hat{N}\rangle=N_{e}$. Substituting the eigenvectors and $\mu$ to the definition of the mean-field parameters, we get a new set of the mean-field parameters, which we denote as $n^{(1)}_{i\sigma}$ ($\sigma=\uparrow, \downarrow$). The above iterative calculations are repeated until the difference $|n^{(m)}_{i\sigma}-n^{(m-1)}_{i\sigma}|$ is smaller than a preseted small positive number $\delta$, for all $i$ and $\sigma$. We then take $n^{(m)}_{i\sigma}$ ($\sigma=\uparrow, \downarrow$) as our convergent mean-field parameters and calculate the eigenstates of the corresponding mean-field model.

We have tested the convergence of the results with respect to the choice of $\delta$ and $n^{(0)}_{i\sigma}$ ($\sigma=\uparrow, \downarrow$). The results in the main text are obtained for $\delta=10^{-5}$. By reducing to $\delta=10^{-8}$, no appreciable changes are found in either the half-filled system or the system with one electron less than half filling. For both the bulk ideal graphane and the real-space graphane lattice with a carbon dimer defect, we take uniform initial states. We therefore have four independent initial mean-field parameters, $n_{1\uparrow}$, $n_{1\downarrow}$, $n_{2\uparrow}$, and $n_{2\downarrow}$. The subindices $1$ and $2$ denote the two sublattices of the carbon atoms. We take $n^{(0)}_{i\sigma}=n_{1\sigma}$ ($\sigma=\uparrow, \downarrow$) if the $i$-th carbon atom belongs to sublattice $1$ and set $n^{(0)}_{i\sigma}=n_{2\sigma}$ ($\sigma=\uparrow, \downarrow$) if the $i$-th carbon atom belongs to sublattice $2$. The charge-ordered ferrimagnetic initial state, which allows the spontaneous emergence of charge ordering and (or) magnetic ordering from the self-consistent calculations, is subject to the constraints $n_{1\uparrow}+n_{1\downarrow}\ne n_{2\uparrow}+n_{2\downarrow}$, $(n_{1\uparrow}-n_{1\downarrow})(n_{2\uparrow}-n_{2\downarrow})<0$, and $(n_{1\uparrow}-n_{1\downarrow})+(n_{2\uparrow}-n_{2\downarrow})\ne0$. We have considered several different sets of initial parameters, including $\{n_{1\uparrow}=0.78, n_{1\downarrow}=0.32, n_{2\uparrow}=0.35, n_{2\downarrow}=0.55\}$, $\{n_{1\uparrow}=0.71, n_{1\downarrow}=0.39, n_{2\uparrow}=0.35, n_{2\downarrow}=0.55\}$, and $\{n_{1\uparrow}=0.57, n_{1\downarrow}=0.45, n_{2\uparrow}=0.45, n_{2\downarrow}=0.53\}$. Consistent results for the convergent mean-field parameters and the eigenspectrum are obtained for both the half-filled system and the system with one electron less than half filling.

\section{Quantum operations}

In this section, we discuss in greater details the quantum gates (i.e., quantum operations) mentioned in the main text. These include the unitary one-qubit gates on the parity qubit of the half-filled system, the unitary one-qubit gates on the spin qubit of the system with one electron less than half filling per carbon dimer, and the nontrivial two-qubit quantum gates on the system with one electron less than half filling per carbon dimer. We will first write down the formula for the most general quantum operations that can be realized in the present system, and then specialize to the cases relevant to our discussions in the main text. In the following constructions for the quantum gates, we assume that the applied stimuli (i.e., the \emph{ac} and the static magnetic fields, and the laser pulse) turn on and turn off abruptly so that we can ignore the time spent for these changes. We also assume that the amplitudes of the external stimuli keep constant during the operations of the quantum gates. We therefore restrict to \emph{rectangular-wave pulses} of the external stimuli. Also notice that, all the constructed quantum gates are assumed to operate on a single carbon dimer defect which is far away from and thus isolated from neighboring carbon dimers.

\subsection{General quantum operations}

The Schr\"{o}dinger equation in the laboratory frame is (we assume $\hbar=1$ hereafter)
\begin{equation}
i\frac{\partial}{\partial t}|\chi(t)\rangle=H|\chi(t)\rangle,
\end{equation}
where $H=H_{C2}+H_{d}$ is the full Hamiltonian. In the basis $\phi^{\dagger}=[a^{\dagger}_{\uparrow},a^{\dagger}_{\downarrow},
b^{\dagger}_{\uparrow},b^{\dagger}_{\downarrow}]$, we have
\begin{eqnarray}
H&=&-\frac{\Delta}{2}\sigma_{z}\otimes s_{0}-\mu_{0}B_{0}\sigma_{0}\otimes s_{z}      \notag \\
&&-\frac{m+m'}{4}\sigma_{0}\otimes s_{z}-\frac{m-m'}{4}\sigma_{z}\otimes s_{z}      \\
&&-\mu_{0}B_{1}\cos(\omega_{1}t)\sigma_{0}\otimes s_{x}-2\gamma\cos(\omega t)\sigma_{y}\otimes s_{0}. \notag
\end{eqnarray}
We introduce the rotating frame defined, in the basis $\phi^{\dagger}$, by \cite{rabi54,nielsenbook}
\begin{equation}
H_{0}=-\frac{\omega}{2}\sigma_{z}\otimes s_{0}-\frac{\omega_{1}}{2}\sigma_{0}\otimes s_{z}.
\end{equation}
The dynamics in the rotating frame is governed by
\begin{equation}
|\varphi(t)\rangle=e^{iH_{0}t}|\chi(t)\rangle,
\end{equation}
and
\begin{equation}
i\frac{\partial}{\partial t}|\varphi(t)\rangle=(e^{iH_{0}t}He^{-iH_{0}t}-H_{0})|\varphi(t)\rangle\equiv\tilde{H}|\varphi(t)\rangle,
\end{equation}
with
\begin{eqnarray}
&&\tilde{H}=-\frac{m-m'}{4}\sigma_{z}\otimes s_{z}-(\frac{m+m'}{4}+\mu_{0}B_{0}-\frac{\omega_{1}}{2})\sigma_{0}\otimes s_{z}   \notag \\
&&-\frac{\mu_{0}B_{1}}{2}\sigma_{0}\otimes s_{x}-\frac{\Delta-\omega}{2}\sigma_{z}\otimes s_{0}-\gamma\sigma_{y}\otimes s_{0}.
\end{eqnarray}
In arriving at the above expression, we have discarded the high-frequency (with angular frequencies $2\omega$ and $2\omega_{1}$) terms, in the spirit of the rotating-wave approximation (RWA) \cite{rabi54}. The RWA, which is valid if $\mu_{0}B_{1}$ ($\gamma$) is much smaller than $\omega_{1}$ ($\omega$) \cite{scheuer14}, is a reasonable approximation for our present system. For rectangular-wave pulses, both during the action of the pulses and when the pulses are turned off, $\tilde{H}$ is time independent. The dynamical evolution of the system is thus simplified. Notice that, the above transformation from the laboratory frame to the rotating frame is formally equivalent to the transformation from the Schr\"{o}dinger picture to the interaction picture. The major difference being that the present $H_{0}$ is not a part of the original model but rather motivated to remove the time dependence of the model.

Now, consider the evolution of an arbitrary state in the Hilbert space of the four in-gap defect states, from $t_{1}$ to $t_{2}$. Formally, we define the evolution operator as
\begin{equation}
|\chi(t_{2})\rangle=U(t_{2},t_{1})|\chi(t_{1})\rangle,
\end{equation}
in the laboratory frame, and
\begin{equation}
|\varphi(t_{2})\rangle=U_{0}(t_{2},t_{1})|\varphi(t_{1})\rangle,
\end{equation}
in the rotating frame. Taking advantage of the definition of Eq.(B4), we can relate the two evolution operators as
\begin{equation}
|\chi(t_{2})\rangle=e^{-iH_{0}t_{2}}U_{0}(t_{2},t_{1})e^{iH_{0}t_{1}}|\chi(t_{1})\rangle=U(t_{2},t_{1})|\chi(t_{1})\rangle,
\end{equation}
Because $\tilde{H}$ is time-independent, we have
\begin{equation}
U_{0}(t_{2},t_{1})=e^{-i\tilde{H}(t_{2}-t_{1})}.
\end{equation}
The full evolution operator in the laboratory frame is thus
\begin{equation}
U(t_{2},t_{1})=e^{-iH_{0}t_{2}}e^{-i\tilde{H}(t_{2}-t_{1})}e^{iH_{0}t_{1}}.
\end{equation}
The initial time of the quantum operation is arbitrary. However, the difference between $|\chi(t_{1})\rangle$ and $|\varphi(t_{1})\rangle$ is irrelevant to the dynamics. We will therefore set $t_{1}=0$, so that at the beginning of the evolution the two wave functions are the same, and the evolution operator depends only on the total time of operation $t$ \cite{vandersypen04}
\begin{equation}
U(t,0)=e^{-iH_{0}t}e^{-i\tilde{H}t}.
\end{equation}

It is necessary to point out that, not all the terms in $H$ are always present in an arbitrary time or time period. For example, in the half-filled system we have $m=m'=0$. Besides, there is no real transition between two states of opposite spin and same parity, because the two states are either both occupied or both empty. Therefore, in discussing general quantum operations applied on the half-filled system, we can set $B_{0}=B_{1}=\omega_{1}=0$. On the other hand, the \emph{ac} magnetic field may also be oriented along a general direction in the $xy$ plane. This extra complication, while more general, is not necessary in the present framework, and therefore we will restrict to the model defined above. With the above point in mind, any quantum operation (i.e., quantum gate) that can be realized in our system is represented as an evolution operator defined by Eq.(B12).

For the system with one electron less than half filling per carbon dimer, only one electron occupies the four in-gap defect states. The Hilbert space is four dimensional, and we denote the four basis as $|j\rangle$ ($j=1,...,4$). The basis $|j\rangle$ ($j=1,...,4$) represents that the $E_{j}$ state is occupied by the electron and all the other three defect states are empty. Starting from the ground state, $|1\rangle$, we can in principle prepare an initial state of the form
\begin{equation}
|\chi(0)\rangle=c_{1}|1\rangle+c_{2}|2\rangle+c_{3}|3\rangle+c_{4}|4\rangle,
\end{equation}
with $c_{j}$ ($j=1,...,4$) general complex numbers that normalize $|\chi(0)\rangle$. This can be achieved by exerting a proper sequence of \emph{ac} magnetic fields and laser pulses. Then, the final state after an evolution (quantum operation) of time $t$ is
\begin{eqnarray}
|\chi(t)\rangle&=&U(t,0)|\chi(0)\rangle=c_{1}U(t,0)|1\rangle+c_{2}U(t,0)|2\rangle    \notag \\
&&+c_{3}U(t,0)|3\rangle+c_{4}U(t,0)|4\rangle.
\end{eqnarray}
For the stoichiometric half-filled system, two electrons occupy the four in-gap defect states. In the ground state, which we denote as $|a\rangle$, $E_{1}$ and $E_{2}$ are both occupied whereas $E_{3}$ and $E_{4}$ are both empty. A resonant laser pulse can connect the ground state to another state $|b\rangle$, in which $E_{1}$ and $E_{2}$ are both empty whereas $E_{3}$ and $E_{4}$ are both occupied. The \emph{ac} magnetic field, which couples $E_{1}$ to $E_{2}$ and $E_{3}$ to $E_{4}$, cannot lead to new states other than $|a\rangle$ and $|b\rangle$. Therefore, the Hilbert space of the half-filled system is two dimensional and have $|a\rangle$ and $|b\rangle$ as the two basis vectors. In this case, we will turn off the ineffective magnetic fields and manipulate the defect states with the laser pulses. Starting from the ground state $|a\rangle$, we can prepare, by applying a proper laser pulse, an initial state of the form
\begin{equation}
|\chi'(0)\rangle=c_{1}'|a\rangle+c_{2}'|b\rangle,
\end{equation}
where $c_{1}'$ and $c_{2}'$ are general complex numbers that normalize $|\chi'(0)\rangle$. The final state after an evolution of time $t$ is obtained by applying $U(t,0)$ on the above state
\begin{equation}
|\chi'(t)\rangle=U(t,0)|\chi'(0)\rangle=c_{1}'U(t,0)|a\rangle+c_{2}'U(t,0)|b\rangle.
\end{equation}

Now, we analyze several general aspects of the evaluation of the evolution operator $U(t,0)$. Because the two terms of $H_{0}$ commute, we have
\begin{eqnarray}
e^{iH_{0}t}&=&e^{-i\frac{\omega t}{2}\sigma_{z}\otimes s_{0}}e^{-i\frac{\omega_{1} t}{2}\sigma_{0}\otimes s_{z}}   \notag \\
&=&(\cos\frac{\omega t}{2}\sigma_{0}\otimes s_{0}-i\sin\frac{\omega t}{2}\sigma_{z}\otimes s_{0})\cdot   \notag \\
&&(\cos\frac{\omega_{1} t}{2}\sigma_{0}\otimes s_{0}-i\sin\frac{\omega_{1} t}{2}\sigma_{0}\otimes s_{z}).
\end{eqnarray}
When there is no \emph{ac} magnetic field (optical field), we set $\omega_{1}=0$ ($\omega=0$), the evolution operator then reduces to a rotation operator about the $z$-axis (we will call it $Z$-rotation in what follows) in the parity (spin) subspace \cite{nielsenbook}.

The $\mu_{0}B_{1}$ and $\gamma$ terms of $\tilde{H}$ are not commutative to the remaining part of $\tilde{H}$. We therefore do not have a similar simple expansion for $\text{exp}(i\tilde{H}t)$ as that for $\text{exp}(iH_{0}t)$. In this case, instead of directly expanding the exponential into its Taylor series, we first diagonalize $\tilde{H}$ with a unitary transformation
\begin{equation}
V^{\dagger}\tilde{H}V=\tilde{H}_{d},
\end{equation}
where $V$ is the desired unitary matrix which diagonalizes $\tilde{H}$, and $\tilde{H}_{d}$ is the diagonal matrix with the eigenvalues of $\tilde{H}$ as the diagonal elements. Then we have
\begin{equation}
e^{i\tilde{H}t}=e^{iV\tilde{H}_{d}V^{\dagger}t}=Ve^{i\tilde{H}_{d}t}V^{\dagger},
\end{equation}
which is easily evaluated after having $V$ and $\tilde{H}_{d}$ in hand.

The solution of $V$ and $\tilde{H}_{d}$ for the most general $\tilde{H}$ is cumbersome. On the other hand, if only one of $\mu_{0}B_{1}$ and $\gamma$ is nonzero, that is if we apply the \emph{ac} magnetic field and laser pulse separately and not simultaneously, then $\tilde{H}$ reduces to the direct sum of two decoupled $2\times2$ subsystems and the corresponding $V$ and $\tilde{H}_{d}$ are easily solvable. We will in what follows focus on this special class of quantum operations and will see that they are sufficient to realize universal quantum computation in the present system. To facilitate the direct-sum decomposition to $\tilde{H}$, we introduce the following projection operators in the spin and parity subspaces
\begin{equation}
P^{s}_{\pm}=\frac{s_{0}\pm s_{z}}{2}, \hspace{1cm} P^{\sigma}_{\pm}=\frac{\sigma_{0}\pm \sigma_{z}}{2}.
\end{equation}
The backward transformations are
\begin{eqnarray}
&&s_{0}=P^{s}_{+}+P^{s}_{-}, \hspace{0.1cm} s_{z}=P^{s}_{+}-P^{s}_{-};   \notag \\ &&\sigma_{0}=P^{\sigma}_{+}+P^{\sigma}_{-}, \hspace{0.1cm} \sigma_{z}=P^{\sigma}_{+}-P^{\sigma}_{-}.
\end{eqnarray}
The projection operators have the following important properties
\begin{eqnarray}
P^{s}_{+}P^{s}_{+}=P^{s}_{+}, \hspace{0.2cm} P^{s}_{-}P^{s}_{-}=P^{s}_{-}, \hspace{0.2cm} P^{s}_{+}P^{s}_{-}=P^{s}_{-}P^{s}_{+}=0;   \notag \\
P^{\sigma}_{+}P^{\sigma}_{+}=P^{\sigma}_{+}, \hspace{0.2cm} P^{\sigma}_{-}P^{\sigma}_{-}=P^{\sigma}_{-}, \hspace{0.2cm} P^{\sigma}_{+}P^{\sigma}_{-}=P^{\sigma}_{-}P^{\sigma}_{+}=0.
\end{eqnarray}

We carry out the direct-sum decompositions to $\tilde{H}$ for the above special cases. First, if neither the \emph{ac} magnetic field nor the laser pulse is applied, we set $\omega_{1}=B_{1}=\omega=\gamma=0$. The remaining terms of $\tilde{H}(\omega_{1}=B_{1}=\omega=\gamma=0)\equiv\tilde{H}_{1}=H_{C2}$ are all mutually commutative, and there is no need to make the direct-sum decomposition. Second, if we apply only the optical field (i.e., laser pulse), we set $\omega_{1}=B_{1}=0$. $B_{0}$ is tunable and may be zero. We decompose $\tilde{H}(\omega_{1}=B_{1}=0)\equiv\tilde{H}_{2}$ as
\begin{equation}
\tilde{H}_{2}=\tilde{H}^{s}_{+}+\tilde{H}^{s}_{-}-(\frac{m+m'}{4}+\mu_{0}B_{0})\sigma_{0}\otimes s_{z},
\end{equation}
where the last term is singled out because it commutes with the remaining part of the model, and
\begin{eqnarray}
\tilde{H}^{s}_{+}&=&
-\frac{m-m'}{4}\sigma_{z}\otimes P^{s}_{+}-\frac{\Delta-\omega}{2}\sigma_{z}\otimes P^{s}_{+}-\gamma\sigma_{y}\otimes P^{s}_{+}      \notag \\
&=&[(-\frac{m-m'}{4}-\frac{\Delta-\omega}{2})\sigma_{z}-\gamma\sigma_{y}]\otimes P^{s}_{+},
\end{eqnarray}
\begin{eqnarray}
\tilde{H}^{s}_{-}&=&
\frac{m-m'}{4}\sigma_{z}\otimes P^{s}_{-}-\frac{\Delta-\omega}{2}\sigma_{z}\otimes P^{s}_{-}-\gamma\sigma_{y}\otimes P^{s}_{-}      \notag \\
&=&[(\frac{m-m'}{4}-\frac{\Delta-\omega}{2})\sigma_{z}-\gamma\sigma_{y}]\otimes P^{s}_{-}.
\end{eqnarray}
Third, if we apply only the \emph{ac} magnetic field, we set $\omega=\gamma=0$. We decompose $\tilde{H}(\omega=\gamma=0)\equiv\tilde{H}_{3}$ as
\begin{equation}
\tilde{H}_{3}=\tilde{H}^{\sigma}_{+}+\tilde{H}^{\sigma}_{-}-\frac{\Delta}{2}\sigma_{z}\otimes s_{0},
\end{equation}
where the last term is singled out because it commutes with the remaining part of the model, and
\begin{equation}
\tilde{H}^{\sigma}_{+}=P^{\sigma}_{+}\otimes[(-\frac{m}{2}-\mu_{0}B_{0}+\frac{\omega_{1}}{2}) s_{z} -\frac{\mu_{0}B_{1}}{2} s_{x}],
\end{equation}
\begin{equation}
\tilde{H}^{\sigma}_{-}
=P^{\sigma}_{-}\otimes[(-\frac{m'}{2}-\mu_{0}B_{0}+\frac{\omega_{1}}{2}) s_{z} -\frac{\mu_{0}B_{1}}{2} s_{x}].
\end{equation}
Because of the properties of the projection operators listed in Eq.(B22), the three terms of $\tilde{H}_{2}$ all commute with each other, as does the three terms of $\tilde{H}_{3}$. The exponentials $\text{exp}(i\tilde{H}_{2,3}t)$ thus reduce to the product of three commutative exponentials that are easy to evaluate.

For the above cases of interest to us, the evolution matrix for the quantum gates can be obtained by applying the following formula
\begin{equation}
R_{\hat{n}}(\theta)=e^{-i\theta\hat{n}\cdot\vec{\sigma}/2}=\cos\frac{\theta}{2}\sigma_{0}-i\sin\frac{\theta}{2}(\hat{n}\cdot\vec{\sigma}),
\end{equation}
where $\hat{n}=(n_{1},n_{2},n_{3})$ is a real unit vector, $\sigma_{0}$ is the $2\times2$ unit matrix, $\vec{\sigma}=(\sigma_{x},\sigma_{y},\sigma_{z})$ with $\sigma_{\alpha}$ ($\alpha=x,y,z$) the Pauli matrices, and the vector inner product $\hat{n}\cdot\vec{\sigma}=n_{1}\sigma_{x}+n_{2}\sigma_{y}+n_{3}\sigma_{z}$. $R_{\hat{n}}(\theta)$ represents a $\theta$-angle rotation of the $\sigma$-spin around the axis along $\hat{n}$, and will be called $N$-rotation (e.g., $X,Y,Z$-rotations) in what follows \cite{nielsenbook}. Notice that, in applying the above formula to the evolution operator expressed in terms of the projection operators, there is a term corresponding to the complementary subspace. As an example, considering the evolution driven by $\tilde{H}^{\sigma}_{+}$, we have
\begin{equation}
e^{-i\tilde{H}^{\sigma}_{+}t}=P^{\sigma}_{+}\otimes[\cos(\Omega t)s_{0}+i\sin(\Omega t)\hat{n}\cdot\vec{s}]+P^{\sigma}_{-}\otimes s_{0},
\end{equation}
where the frequency
\begin{equation}
\Omega=\sqrt{(\frac{\omega_{1}-m}{2}-\mu_{0}B_{0})^{2}+(\frac{\mu_{0}B_{1}}{2})^{2}},
\end{equation}
and the unit vector along the rotation axis
\begin{equation}
\hat{n}=(n_{1},n_{2},n_{3})=\frac{1}{\Omega}(\frac{\mu_{0}B_{1}}{2},0,\frac{m-\omega_{1}}{2}+\mu_{0}B_{0}).
\end{equation}
The last term, $P^{\sigma}_{-}\otimes s_{0}$, which is a unit operator in the complementary subspace, should be included in the expansion to arrive at the correct final result. When $\omega_{1}=m+2\mu_{0}B_{0}$, the $\tilde{H}^{\sigma}_{+}$ subspace is at resonance. Assuming $B_{1}>0$, we have $\Omega=\mu_{0}B_{1}/2$ and $\hat{n}=(1,0,0)$. When the system is evolved by a time $t=(\frac{\pi}{2})/\Omega=\pi/(\mu_{0}B_{1})$, the two basis states of the $\tilde{H}^{\sigma}_{+}$ subspace (i.e., $|1\rangle$ and $|2\rangle$) are exchanged (the Rabi flopping \cite{rabi54,nielsenbook}). On the other hand, when the complementary subspace associated with $\tilde{H}^{\sigma}_{-}$ is at resonance, we have $\omega_{1}=m'+2\mu_{0}B_{0}$. The $\tilde{H}^{\sigma}_{+}$ subspace is now off resonance, with $\Omega=\sqrt{(m-m')^{2}+(\mu_{0}B_{1})^{2}}/2$ and $\hat{n}=(n_{1},n_{2},n_{3})=(\mu_{0}B_{1},0,m-m')/(2\Omega)$. Usually, we have $|m-m'|\gg|\mu_{0}B_{1}|$. So we have $|n_{3}|\gg|n_{1}|$, and the hybridization between $|1\rangle$ and $|2\rangle$ is negligible for arbitrary evolution time.

\subsection{One-qubit gates on the parity qubit of the half-filled system}

We first consider the quantum operations that can be applied to the half-filled system. From the main text, we have $m=m'=0$ in this case. The four in-gap defect states gather into two two-fold degenerate groups. In the presence of a static magnetic field $B_{0}$, the two-fold spin degeneracies of the two groups are broken to the same extent. If the Zeeman interaction from $B_{0}$ does not change the occupancy of the system, the \emph{ac} magnetic field cannot trigger a real transition, because the two relevant states are always either both occupied or both empty. Therefore, in the half-filled system, the spin degrees of freedom is inert and can be ignored from a practical point of view. We thus set $\omega_{1}=B_{1}=B_{0}=0$ for the half-filled system. The system is now effectively a parity qubit, which is tunable by the optical field through the electric dipole transition.

The model reduces to
\begin{equation}
H_{0}=-\frac{\omega}{2}\sigma_{z}\otimes s_{0},
\end{equation}
and
\begin{equation}
\tilde{H}=-\frac{\Delta-\omega}{2}\sigma_{z}\otimes s_{0}-\gamma\sigma_{y}\otimes s_{0}.
\end{equation}
At resonance, $\omega=\Delta$, and the first term of $\tilde{H}$ vanishes. An evolution of duration $t$ is thus \begin{eqnarray}
U(t,0)&=&e^{-iH_{0}t}e^{-i\tilde{H}t}=e^{i\frac{\Delta t}{2}\sigma_{z}\otimes s_{0}}e^{i\gamma t\sigma_{y}\otimes s_{0}}      \notag \\
&=&R^{\sigma}_{z}(-\Delta t)R^{\sigma}_{y}(-2\gamma t),
\end{eqnarray}
which is formally a composite gate of a $Y$-rotation and a $Z$-rotation in the parity subspace \cite{nielsenbook}. By tuning the evolution time $t$, the strength of the optical field $\gamma$, and by exerting a sequential multipulse operation, we can achieve arbitrary unitary transformations (i.e., quantum gates) on the parity qubit. What follows we construct the Hadamard gate (H), the phase gate (S), and the $\pi/8$ gate (T). It is well known that these three gates can be compounded to approximate arbitrary single-qubit gates \cite{nielsenbook}.

In the parity subspace, and suppressing the inactive spin degrees of freedom, the three gates can be represented in terms of the $Y$-rotations and $Z$-rotations as
\begin{eqnarray}
&&S=\begin{pmatrix} 1 & 0 \\ 0 & i \end{pmatrix}=e^{i\frac{\pi}{4}}R_{z}(\frac{\pi}{2}),      \notag \\
&&T=\begin{pmatrix} 1 & 0 \\ 0 & e^{i\frac{\pi}{4}} \end{pmatrix}=e^{i\frac{\pi}{8}}R_{z}(\frac{\pi}{4}),   \notag \\
&&H=\frac{1}{\sqrt{2}}\begin{pmatrix} 1 & 1 \\ 1 & -1 \end{pmatrix}=e^{i\frac{\pi}{2}}R_{y}(\frac{\pi}{2})R_{z}(\pi).
\end{eqnarray}
The global phase factors in front of the final expressions have no observable effects and can be ignored. The phase (S) and $\pi/8$ (T) gates are thus realized by turning off the laser pulse (i.e., $\gamma=0$) and freely evolve the system by a time of $7\pi/(2\Delta)$ and $15\pi/(4\Delta)$, respectively. Note that, up to the $2\pi$ phase ambiguity, we have $-\pi/4\equiv7\pi/4$ and $-\pi/8\equiv15\pi/8$.

For the Hadamard (H) gate, we firstly freely evolve the system by a time of $3\pi/\Delta$ to get $R_{z}(\pi)$. Then we apply the laser pulse. If we can continuously tune $\gamma$, we can choose $\gamma$ and the time duration $t$, such that $\Delta t=4N\pi$ and $4\gamma t=7\pi$ are satisfied at the same time, and get $R_{y}(\frac{\pi}{2})$. Here $N$ is a large positive integer since $\Delta\gg\gamma$ applies. If $\gamma$ is not easily tunable, we can still apply the laser pulse for a time of $4\gamma t=7\pi$ and then we freely evolve the system for an additional time of $t'$, such that $\Delta(t+t')=4N\pi$ with $N$ a positive integer.

\subsection{One-qubit gates on the spin qubit of the system with one electron less than half filling per carbon dimer}

In the system with one electron less than half filling, we can have different ways of encoding a single qubit. One is to combine $|1\rangle$ and $|2\rangle$ with a resonant \emph{ac} magnetic field, another is by combining $|1\rangle$ and $|3\rangle$ with a resonant laser pulse. Here, we focus on the first approach, because it utilizes the two lowest energy levels and is the most natural encoding of a qubit in this system. The second approach can be realized in a similar manner, which is formally equivalent to the discussions in the previous section, if we restrict to the subspace of $|1\rangle$ and $|3\rangle$ and ignore the evolution of the subspace consisting of $|2\rangle$ and $|4\rangle$.

The full model for the quantum gates is $\tilde{H}_{3}$, together with the model for the rotating frame
\begin{equation}
H_{0}=-\frac{\omega_{1}}{2}\sigma_{0}\otimes s_{z}=-\frac{\omega_{1}}{2}(P^{\sigma}_{+}+P^{\sigma}_{-})\otimes s_{z}.
\end{equation}
The considered spin qubit is described by $\tilde{H}^{\sigma}_{+}$ in the rotating frame. The subsystem described by $\tilde{H}^{\sigma}_{-}$ in the rotating frame is decoupled from the above spin qubit. We consider them together in this section to get a complete picture about the evolution of the system consisting of all the four defect states, which is relevant to the two-qubit gates to be discussed in the next section.

The quantum evolution operator for a duration of $t$ is
\begin{equation}
U(t,0)=e^{-iH_{0}t}e^{-i\tilde{H}_{3}t}=U^{\sigma}_{+}(t,0)U^{\sigma}_{-}(t,0),
\end{equation}
where
\begin{equation}
U^{\sigma}_{+}(t,0)=e^{i\frac{1}{2}\omega_{1}t P^{\sigma}_{+}\otimes s_{z}}e^{-i \tilde{H}^{\sigma}_{+}t} e^{i\frac{1}{2}\Delta tP^{\sigma}_{+}\otimes s_{0}},
\end{equation}
\begin{equation}
U^{\sigma}_{-}(t,0)=e^{i\frac{1}{2}\omega_{1}t P^{\sigma}_{-}\otimes s_{z}}e^{-i \tilde{H}^{\sigma}_{-}t} e^{-i\frac{1}{2}\Delta tP^{\sigma}_{-}\otimes s_{0}}.
\end{equation}
Since the $\tilde{H}^{\sigma}_{+}$ subsystem is taken to encode the qubit, we set it to be at resonance by setting $\omega_{1}=m+2\mu_{0}B_{0}\equiv\tilde{m}$. We thus have
\begin{equation}
U^{\sigma}_{+}(t,0)=e^{i\frac{1}{2}\tilde{m}t P^{\sigma}_{+}\otimes s_{z}}e^{i\frac{1}{2}\mu_{0}B_{1}t P^{\sigma}_{+}\otimes s_{x}}e^{i\frac{1}{2}\Delta tP^{\sigma}_{+}\otimes s_{0}},
\end{equation}
\begin{eqnarray}
&&U^{\sigma}_{-}(t,0)=e^{i\frac{1}{2}\tilde{m}t P^{\sigma}_{-}\otimes s_{z}}e^{-i\frac{1}{2}(m-m')t P^{\sigma}_{-}\otimes s_{z}+i\frac{1}{2}\mu_{0}B_{1}t P^{\sigma}_{-}\otimes s_{x}}\cdot    \notag \\
&&\cdot e^{-i\frac{1}{2}\Delta tP^{\sigma}_{-}\otimes s_{0}}=e^{i\frac{1}{2}\tilde{m}t P^{\sigma}_{-}\otimes s_{z}}e^{i\Omega t P^{\sigma}_{-}\otimes\hat{n}\cdot\vec{s}}e^{-i\frac{1}{2}\Delta tP^{\sigma}_{-}\otimes s_{0}},
\end{eqnarray}
where we have defined the frequency $\Omega$ and unit vector $\hat{n}$ as
\begin{equation}
\Omega=\sqrt{(\frac{m-m'}{2})^{2}+(\frac{\mu_{0}B_{1}}{2})^{2}},
\end{equation}
and
\begin{equation}
\hat{n}=(n_{1},n_{2},n_{3})=(\frac{\mu_{0}B_{1}}{2\Omega},0,\frac{m'-m}{2\Omega}).
\end{equation}
In the subspace of $\tilde{H}^{\sigma}_{+}$, by tuning the evolution time $t$, the strength of the \emph{ac} magnetic field $B_{1}$ and the static magnetic field $B_{0}$, and by exerting a sequential multipulse operation, we can achieve arbitrary unitary transformations (i.e., quantum gates) on the spin qubit. The last factor of  $\tilde{H}^{\sigma}_{+}$, which contributes a global phase factor to the evolution operator, can be neglected if we restrict to this subspace. We will consider the phase (S) gate, the $\pi/8$ (T) gate, and the Hadamard (H) gate in the subspace associated with $\tilde{H}^{\sigma}_{+}$, and see whether we can leave the subspace associated with $\tilde{H}^{\sigma}_{-}$ completely unaltered at the same time.

For the S and T gates on the $\tilde{H}^{\sigma}_{+}$ subspace, we set $B_{1}=0$ and freely evolve the spin qubit by
\begin{equation}
U^{\sigma}_{+}(t,0)=e^{i\frac{1}{2}\tilde{m}tP^{\sigma}_{+}\otimes s_{z}}e^{i\frac{1}{2}\Delta tP^{\sigma}_{+}\otimes s_{0}}.
\end{equation}
The corresponding evolution in the subspace of $\tilde{H}^{\sigma}_{-}$ is driven by
\begin{equation}
U^{\sigma}_{-}(t,0)=e^{i\frac{1}{2}\tilde{m}'tP^{\sigma}_{-}\otimes s_{z}}e^{-i\frac{1}{2}\Delta tP^{\sigma}_{-}\otimes s_{0}},
\end{equation}
where $\tilde{m}'=m'+2\mu_{0}B_{0}=\tilde{m}+(m'-m)$.

To realize an S gate on the $\tilde{H}^{\sigma}_{+}$ subspace by a free evolution of time $t_{1}$, and at the same time keep the $\tilde{H}^{\sigma}_{-}$ subspace completely inert, we require
\begin{equation}
\frac{\tilde{m}t_{1}}{2}=2N_{1}\pi+\frac{7\pi}{4}, \hspace{0.5cm}
\frac{\tilde{m}'t_{1}}{2}=2N_{2}\pi, \hspace{0.5cm}
\Delta t_{1}=2N_{3}\pi,
\end{equation}
where $N_{1}$, $N_{2}$, and $N_{3}$ are arbitrary positive integers. The three equalities should be fulfilled at the same time. This constraint can be satisfied by the following procedure. Firstly, we tune the magnitude of $B_{0}$ (which is the only tunable model parameter) so that
\begin{equation}
\frac{(\tilde{m}'/2)}{\Delta}=\frac{N_{4}}{N_{5}},
\end{equation}
where $N_{4}$ and $N_{5}$ are positive integers that are mutually prime and as small as possible. Then we define $\omega_{0}$ as a common divisor of $\tilde{m}'/2$ and $\Delta$
\begin{equation}
\omega_{0}=\frac{\tilde{m}'}{2N_{4}}=\frac{\Delta}{N_{5}}.
\end{equation}
Then, we search for two positive integers $N_{1}$ and $N_{6}$, such that we can approximately have
\begin{equation}
\frac{2\tilde{m}}{\omega_{0}}=\frac{8N_{1}+7}{2N_{6}},
\end{equation}
up to a certain preseted precision. Finally, we determine $t_{1}$ by requiring
\begin{equation}
\frac{\tilde{m}}{2}t_{1}=2N_{1}\pi+\frac{7\pi}{4}=\frac{(8N_{1}+7)\pi}{4}.
\end{equation}
For this $t_{1}$ we have
\begin{equation}
\omega_{0}t_{1}=2N_{6}\pi.
\end{equation}
And therefore
\begin{eqnarray}
&&\frac{\tilde{m}'}{2}t_{1}=N_{4}\omega_{0}t_{1}=2N_{4}N_{6}\pi\equiv2N_{2}\pi,    \notag \\
&&\Delta t_{1}=N_{5}\omega_{0}t_{1}=2N_{5}N_{6}\pi\equiv2N_{3}\pi.
\end{eqnarray}
In this manner, by tuning $B_{0}$ and $t_{1}$, we can realize a pure phase (S) gate on the chosen spin qubit and leave the complementary subspace completely unaltered. The accuracy of this gate is determined by the preseted precision for determining $N_{1}$ and $N_{6}$.

A pure T gate in the subspace of $\tilde{H}^{\sigma}_{+}$ can be constructed in the same manner. Suppose the T gate is realized by a free evolution of time $t_{2}$. To keep the $\tilde{H}^{\sigma}_{-}$ subspace completely inert, we require
\begin{equation}
\frac{\tilde{m}t_{2}}{2}=2N_{1}'\pi+\frac{15\pi}{8}, \hspace{0.5cm}
\frac{\tilde{m}'t_{2}}{2}=2N_{2}'\pi, \hspace{0.5cm}
\Delta t_{2}=2N_{3}'\pi,
\end{equation}
where $N_{1}'$, $N_{2}'$, and $N_{3}'$ are positive integers to be determined. We follow the same procedure for the S gate to fulfill the above requirements. In the first step, we tune the magnitude of $B_{0}$ to make
\begin{equation}
\frac{(\tilde{m}'/2)}{\Delta}=\frac{N_{4}'}{N_{5}'}.
\end{equation}
Since this is the same condition as the above one, we take $N_{4}'=N_{4}$ and $N_{5}'=N_{5}$, and define $\omega_{0}$ in terms of Eq.(B49). Then, we search for two positive integers $N_{1}'$ and $N_{6}'$, such that we can approximately have
\begin{equation}
\frac{4\tilde{m}}{\omega_{0}}=\frac{16N_{1}'+15}{2N_{6}'},
\end{equation}
up to a preseted precision. Finally, we determine $t_{2}$ by requiring
\begin{equation}
\frac{\tilde{m}}{2}t_{2}=2N_{1}'\pi+\frac{15\pi}{8}=\frac{(16N_{1}'+15)\pi}{8}.
\end{equation}
For this $t_{2}$ we have
\begin{equation}
\omega_{0}t_{2}=2N_{6}'\pi.
\end{equation}
And therefore
\begin{eqnarray}
&&\frac{\tilde{m}'}{2}t_{2}=N_{4}\omega_{0}t_{1}=2N_{4}N_{6}'\pi\equiv2N_{2}'\pi,     \notag \\
&&\Delta t_{2}=N_{5}\omega_{0}t_{2}=2N_{5}N_{6}'\pi\equiv2N_{3}'\pi.
\end{eqnarray}
Therefore, we can realize a pure $\pi/8$ (T) gate on the chosen spin qubit and leave the complementary subspace completely unaltered. The accuracy of this gate is determined by the preseted precision for determining $N_{1}'$ and $N_{6}'$.

We next construct the Hadamard (H) gate on the subspace of $\tilde{H}^{\sigma}_{+}$. Again, we want to keep the $\tilde{H}^{\sigma}_{-}$ subspace completely inert at the end of this operation. In terms of the $X$-rotations and $Z$-rotations in the space of the target spin qubit, the H gate can be realized through
\begin{equation}
H=e^{i\frac{\pi}{2}}R_{z}(\frac{\pi}{2})R_{x}(\frac{\pi}{2})R_{z}(\frac{\pi}{2}).
\end{equation}
This realization of the H gate is related to that in the previous section through the identity
\begin{equation}
R_{y}(\theta)=R_{z}(\frac{\pi}{2})R_{x}(\theta)R_{z}(-\frac{\pi}{2}).
\end{equation}
The H gate is therefore realized through three consecutive operations. The analysis is simplified by noticing from Eq.(B36) that $R_{z}(\frac{\pi}{2})$ is equivalent to the phase (S) gate up to a global phase factor, and we can rewrite
\begin{equation}
H=e^{i\frac{\pi}{2}}R_{z}(\frac{\pi}{2})R_{x}(\frac{\pi}{2})R_{z}(\frac{\pi}{2})=SR_{x}(\frac{\pi}{2})S.
\end{equation}
By applying the two phase (S) gates in the same manner as that defined above, the construction of the H gate reduces to the design of a quantum gate realizing $R_{x}(\frac{\pi}{2})$. Referring to the evolution operators defined by Eqs.(B41) and (B42), the desired $R_{x}(\frac{\pi}{2})$ gate in the subspace of $\tilde{H}^{\sigma}_{+}$ is realized by imposing the following constraints
\begin{eqnarray}
&&\frac{\mu_{0}B_{1}t_{3}}{2}=2N_{1}''\pi+\frac{7\pi}{4}, \hspace{0.25cm}
\frac{\tilde{m}t_{3}}{2}=2N_{2}''\pi, \hspace{0.25cm}    \notag \\
&&\Omega t_{3}=2N_{3}''\pi, \hspace{0.25cm}
\Delta t_{3}=2N_{4}''\pi,
\end{eqnarray}
where $N_{i}''$ ($i=1,...,4$) are a set of positive integers.
By the following procedure, we can find the parameters that fulfill the above conditions. Firstly, tuning the magnitudes of $B_{0}$ and $B_{1}$ to make the three relevant frequencies in integral proportions
\begin{equation}
(\tilde{m}/2):\Omega:\Delta=N_{5}'':N_{6}'':N_{7}'',
\end{equation}
where $N_{5}''$, $N_{6}''$, and $N_{7}''$ are integers that are as small as possible. In particular, the three integers have no common divisor larger than 1. Now define the common divisor $\omega_{0}''$ of the three frequencies as
\begin{equation}
\omega_{0}''=\frac{\tilde{m}}{2N_{5}''}=\frac{\Omega}{N_{6}''}=\frac{\Delta}{N_{7}''}.
\end{equation}
Next, we search for two positive integers $N_{1}''$ and $N_{8}''$, that are small enough and can satisfy the following condition
\begin{equation}
\frac{2\mu_{0}B_{1}}{\omega_{0}''}=\frac{8N_{1}''+7}{2N_{8}''},
\end{equation}
within a certain precision. Finally, we determine the time $t_{3}$ by requiring
\begin{equation}
\frac{\mu_{0}B_{1}t_{3}}{2}=2N_{1}''\pi+\frac{7\pi}{4}=\frac{8N_{1}''+7}{4}\pi.
\end{equation}
For this $t_{3}$, we have
\begin{equation}
\omega_{0}''t_{3}=2N_{8}''\pi.
\end{equation}
The last three conditions in Eq.(B63) are thus fulfilled as
\begin{eqnarray}
&&\frac{\tilde{m}t_{3}}{2}=N_{5}''\omega_{0}''t_{3}=2N_{5}''N_{8}''\pi\equiv2N_{2}''\pi,
\hspace{0.25cm}     \notag \\
&&\Omega t_{3}=N_{6}''\omega_{0}''t_{3}=2N_{6}''N_{8}''\pi\equiv2N_{3}''\pi,
\hspace{0.25cm}     \notag \\
&&\Delta t_{3}=N_{7}''\omega_{0}''t_{3}=2N_{7}''N_{8}''\pi\equiv2N_{4}''\pi.
\end{eqnarray}
By adopting the above parameters, we can realize the $R_{x}(\frac{\pi}{2})$ gate in the subspace of $\tilde{H}^{\sigma}_{+}$, and at the same time keep the subspace of $\tilde{H}^{\sigma}_{-}$ completely unaltered. Compounding this gate in between two S gates defined above, we can realize the desired H gate in the subspace of $\tilde{H}^{\sigma}_{+}$, and at the same time keep the subspace of $\tilde{H}^{\sigma}_{-}$ completely unaltered. We note by passing that, in searching for the parameters fulfilling the four constraints in Eq.(B63), we have three free parameters ($B_{0}$, $B_{1}$, $t_{3}$). This is to be compared to the cases for the  S and T gates, where we have two free parameters to meet the requirements imposed by three equalities.

\subsection{Two-qubit gates in the system with one electron less than half filling per carbon dimer}

A single nontrivial two-qubit gate, together with arbitrary one-qubit gates, are known to be able to represent any nontrivial two-qubit gates \cite{nielsenbook,barenco95,deutsch95,lloyd95,bremner02}. We thus focus on one typical two-qubit gate, the controlled-NOT (CNOT) gate.

As was pointed out in the main text, a two-qubit system can in principle be encoded in arbitrary four states, once they can be manipulated in a nontrivial manner at the two-qubit level \cite{nielsenbook,kessel99,kessel02,kiktenko15}. One type of manipulations are realized by applying appropriate external stimuli, such as the laser pulses and the \emph{ac} magnetic fields considered in the present work. A simplest encoding of the two-qubit system in the present system, without invoking the inter-dimer coupling, is in terms of the four in-gap energy levels associated with a single carbon dimer defect with one electron less than half filling.

There can be two different encodings of the two-qubit system. In the first encoding, we take the two spins as the target qubit and the two parities as the control qubit. In the second encoding, we take the two parities as the target qubit and the two spins as the control qubit. We will focus on the first encoding. In this encoding, we have the following designations
\begin{equation}
|1\rangle\equiv|00\rangle, \hspace{0.25cm} |2\rangle\equiv|01\rangle, \hspace{0.25cm} |3\rangle\equiv|10\rangle, \hspace{0.25cm} |4\rangle\equiv|11\rangle,
\end{equation}
where in the conventional expression of the two-qubit state, $|mn\rangle$, the first qubit is in the $m$-th  ($m=0,1$) state and the second qubit is in the $n$-th ($n=0,1$) state. The above correspondence can be made more exact by referring to a model of two spin-$1/2$ qubits coupled through the Ising spin coupling
\begin{equation}
H_{spin}=-AZ_{1}-BZ_{2}+JZ_{1}Z_{2}.
\end{equation}
In the convention that $0$ represents spin up ($Z=+1$) and $1$ represents spin down ($Z=-1$), the eigen-spectrum of $H_{spin}$ is
\begin{eqnarray}
&&E_{00}=-A-B+J, \hspace{0.25cm} E_{01}=-A+B-J, \hspace{0.25cm}   \notag \\
&&E_{10}=A-B-J, \hspace{0.25cm} E_{11}=A+B+J.
\end{eqnarray}
In the absence of any external fields (i.e., $B_{0}=B_{1}=\gamma=0$), the in-gap defect states associated with a single isolated carbon dimer are governed by $H_{C2}$, which has an eigen-spectrum
\begin{eqnarray}
&&E_{1}=-\frac{\Delta}{2}-\frac{m}{2}, \hspace{0.25cm} E_{2}=-\frac{\Delta}{2}+\frac{m}{2}, \hspace{0.25cm}  \notag \\
&&E_{3}=\frac{\Delta}{2}-\frac{m'}{2}, \hspace{0.25cm} E_{4}=\frac{\Delta}{2}+\frac{m'}{2}.
\end{eqnarray}
The two spectra are equivalent if we make the following identifications
\begin{equation}
A=\frac{\Delta}{2}, \hspace{0.5cm} B=\frac{m+m'}{4}, \hspace{0.5cm} J=\frac{m'-m}{4}.
\end{equation}
This formal equivalence is even clearer, if we recall the expression of $H_{C2}$ in the basis of $\phi^{\dagger}$,
\begin{equation}
H_{C2}=-\frac{\Delta}{2}\sigma_{z}\otimes s_{0}-\frac{m+m'}{4}\sigma_{0}\otimes s_{z}+\frac{m'-m}{4}\sigma_{z}\otimes s_{z}.
\end{equation}

In the basis $\phi^{\dagger}$, the CNOT gate is represented as
\begin{equation}
\text{CNOT}= \begin{pmatrix} 1 & 0 & 0 & 0 \\ 0 & 1 & 0 & 0 \\  0 & 0 & 0 & 1 \\  0 & 0 & 1 & 0 \end{pmatrix} =P^{\sigma}_{+}\otimes s_{0}+P^{\sigma}_{-}\otimes s_{x}.
\end{equation}
To realize the NOT gate in the subspace associated with $\tilde{H}^{\sigma}_{-}$, we have to set this subsystem at resonance and so that the subsystem associated with $\tilde{H}^{\sigma}_{+}$ is off resonance. In this sense, the present situation is complementary to the $R_{x}(\pi/2)$ gate considered in the previous section. At resonance, we have $\omega_{1}=\tilde{m}'=m'+2\mu_{0}B_{0}=\tilde{m}+(m'-m)$. The evolution operators in the two independent subspaces are now
\begin{eqnarray}
&&U^{\sigma}_{+}(t,0)=e^{i\frac{1}{2}\tilde{m}'t P^{\sigma}_{+}\otimes s_{z}}e^{i\frac{1}{2}(m-m')t P^{\sigma}_{+}\otimes s_{z}+i\frac{1}{2}\mu_{0}B_{1}t P^{\sigma}_{+}\otimes s_{x}}\cdot      \notag \\
&&\cdot e^{i\frac{1}{2}\Delta tP^{\sigma}_{+}\otimes s_{0}}=e^{i\frac{1}{2}\tilde{m}'t P^{\sigma}_{+}\otimes s_{z}}e^{i\Omega t P^{\sigma}_{+}\otimes\hat{n}\cdot\vec{s}}e^{i\frac{1}{2}\Delta tP^{\sigma}_{+}\otimes s_{0}},
\end{eqnarray}
\begin{equation}
U^{\sigma}_{-}(t,0)=e^{i\frac{1}{2}\tilde{m}'t P^{\sigma}_{-}\otimes s_{z}}e^{i\frac{1}{2}\mu_{0}B_{1}t P^{\sigma}_{-}\otimes s_{x}}e^{-i\frac{1}{2}\Delta tP^{\sigma}_{-}\otimes s_{0}},
\end{equation}
where the frequency and the unit vector are defined as
\begin{equation}
\Omega=\sqrt{(\frac{m-m'}{2})^{2}+(\frac{\mu_{0}B_{1}}{2})^{2}},
\end{equation}
\begin{equation}
\hat{n}=(n_{1},n_{2},n_{3})=(\frac{\mu_{0}B_{0}}{2\Omega},0,\frac{m-m'}{2\Omega}).
\end{equation}

To realize the desired CNOT gate, we should realize a NOT gate on the odd-parity subspace with $U^{\sigma}_{-}(t,0)$ and at the same time keep the even-parity subspace unaltered. We free the time labels and integral number labels used in the previous section. The above purpose can be realized by a single evolution of a finite time duration $t$ in the presence of the \emph{ac} magnetic field ($\omega=\gamma=0$), if the following conditions are simultaneously fulfilled
\begin{eqnarray}
&&\frac{\mu_{0}B_{1}t}{2}=2N_{1}\pi+\frac{\pi}{2},
\hspace{0.5cm}   \frac{\tilde{m}'t}{2}=2N_{2}\pi,  \hspace{0.5cm}    \notag \\
&&\Omega t=2N_{3}\pi,
\hspace{0.5cm}   \Delta t=2N_{4}\pi.
\end{eqnarray}
By a procedure closely parallel to that adopted in the previous section to construct the H gate, we can find the parameters that satisfy the above conditions. Firstly, tune the magnitudes of $B_{0}$ and $B_{1}$ so that the three relevant frequencies are in integral proportions
\begin{equation}
(\tilde{m}'/2):\Omega:\Delta=N_{5}:N_{6}:N_{7},
\end{equation}
where $N_{5}$, $N_{6}$, and $N_{7}$ are positive integers that are as small as possible. In particular, the three integers have no common divisor larger than 1. Now define the common divisor $\omega_{0}$ of the three frequencies as
\begin{equation}
\omega_{0}=\frac{\tilde{m}'}{2N_{5}}=\frac{\Omega}{N_{6}}=\frac{\Delta}{N_{7}}.
\end{equation}
Next, we search for two positive integers $N_{1}$ and $N_{8}$, that are small enough and can satisfy the following condition
\begin{equation}
\frac{\mu_{0}B_{1}}{\omega_{0}}=\frac{4N_{1}+1}{2N_{8}},
\end{equation}
within a preseted precision. Finally, we determine the time $t$ by requiring
\begin{equation}
\frac{\mu_{0}B_{1}t}{2}=2N_{1}\pi+\frac{\pi}{2}=\frac{4N_{1}+1}{2}\pi.
\end{equation}
For this $t$, we have
\begin{equation}
\omega_{0}t=2N_{8}\pi.
\end{equation}
The last three conditions in Eq.(B81) are thus fulfilled as
\begin{eqnarray}
&&\frac{\tilde{m}'t}{2}=N_{5}\omega_{0}t=2N_{5}N_{8}\pi\equiv2N_{2}\pi,
\hspace{0.5cm}  \notag \\
&&\Omega t=N_{6}\omega_{0}t=2N_{6}N_{8}\pi\equiv2N_{3}\pi,
\hspace{0.5cm}   \notag \\
&&\Delta t=N_{7}\omega_{0}t=2N_{7}N_{8}\pi\equiv2N_{4}\pi.
\end{eqnarray}
In the above manner, we can realize the CNOT gate on the two-qubit system encoded in the four nondegenerate in-gap defect states associated with a single carbon dimer with one electron less than stoichiometry.

\end{appendix}


\end{document}